\documentclass[12pt,reqno,notitlepage]{amsart}

\usepackage[letterpaper,margin=1in]{geometry}

\usepackage{anyfontsize}
\usepackage[T1]{fontenc}

\usepackage{amsmath,amssymb,amsthm}

\usepackage{graphicx}
\usepackage[svgnames]{xcolor}
\usepackage{pgf}
\usepackage{tikz}
\usepackage{mathrsfs}
\usepackage{stmaryrd}
\SetSymbolFont{stmry}{bold}{U}{stmry}{m}{n}
\usepackage{layout}
\usepackage{mathtools}
\usepackage{paralist}
\usepackage{multirow}
\usepackage{comment}
\usepackage{rotating}
\usepackage{chngcntr}
\usepackage{apptools}

\usepackage{mathpazo}

\usepackage{xspace}
\usepackage{bm}
\usepackage{setspace}
\usepackage{accents}
\usepackage{enumitem}
\usepackage{footnote}
\usepackage{tabularx}
\usepackage{threeparttable}

\makesavenoteenv{tabular}
\makesavenoteenv{table}

\usepackage[colorlinks=true, linkcolor=Maroon, urlcolor=Maroon, citecolor=Maroon]{hyperref}

\usepackage[authoryear]{natbib}

\usepackage{makecell}

\theoremstyle{plain}
\newtheorem{assumption}{Assumption}

\newtheorem{theorem}{Theorem}
\newtheorem{corollary}{Corollary}

\theoremstyle{remark}

\newtheorem{remark}{Remark}

\usepackage{todonotes}

\renewcommand{\Pr}{\mathbb{P}}

\newcommand{\Exp}{\mathbb{E}}

\newcommand{\convd}{\stackrel{d}{\rightarrow}}

\usepackage{subfig}
\newcommand{\one}{\mathbb{1}}

\newcommand{\att}{\textrm{ATT}}

\newcommand{\ES}{\textrm{ES}}
\newcommand{\AP}{\textrm{AP}}
\newcommand{\NP}{\textrm{NP}}
\newcommand{\TP}{\textrm{TP}}
\newcommand{\TC}{\textrm{TC}}

\usepackage{centernot}

\usepackage{cleveref}
\crefname{figure}{figure}{figures}
\creflabelformat{figure}{#2#1#3}

\crefname{equation}{equation}{equations}
\crefrangeformat{equation}{(#3#1#4) to~(#5#2#6)}
\creflabelformat{equation}{(#2#1#3)}

\crefname{lemma}{lemma}{lemmas}
\creflabelformat{lemma}{#2#1#3}
\crefname{proposition}{proposition}{propositions}
\creflabelformat{proposition}{#2#1#3}

\crefname{corollary}{corollary}{corollaries}
\creflabelformat{corollary}{#2#1#3}

\crefname{condition}{condition}{conditions}
\creflabelformat{condition}{#2#1#3}
\crefname{assumption}{assumption}{assumptions}
\creflabelformat{assumption}{#2#1#3}
\crefname{remark}{remark}{remarks}
\creflabelformat{remark}{#2#1#3}
\crefname{appendix}{appendix}{appendices}
\creflabelformat{appendix}{#2#1#3}

\allowdisplaybreaks
\begin{document}

\renewcommand{\thefootnote}{\fnsymbol{footnote}}

\begin{center}
\Large\textsc{Learning the Effect of Persuasion via Difference-In-Differences}\footnote{We would like to thank Peng Ding, Stefano DellaVigna, Vitor Possebom, Jonathan Roth, Pedro Sant'Anna, Zeyang (Arthur) Yu, and the seminar participants at
McMaster University,
Munich Econometrics Seminar,
Seoul National University,
Sungkyunkwan University,
the University of Kentucky,
the University of Toronto,
and the 2025 World Congress of the Econometric Society
for encouraging and helpful comments.}
\vspace{1ex}
\end{center}

\vspace{.3cm}
\begin{center}
    \begin{tabular}{ccc}
        \large\textsc{Sung Jae Jun}\footnote{Department of Economics, suj14@psu.edu.}
        &
        &
        \large\textsc{Sokbae Lee}\footnote{(Corresponding Author) Department of Economics, sl3841@columbia.edu. }
        \\
        Pennsylvania State University
        &
        &
        Columbia University
    \end{tabular}
\end{center}

\begin{center}
\vspace{0.3cm}
    August 27, 2026
\vspace{0.3cm}
\end{center}

\noindent
\textbf{Abstract.}
We develop a difference-in-differences framework to measure the persuasive impact of informational treatments on behavior in staggered treatment settings. We introduce two causal parameters, the forward and backward average persuasion rates on the treated, which refine the average treatment effect on the treated. The forward rate excludes cases of ``preaching to the converted,'' while the backward rate omits ``talking to a brick wall'' cases. The backward rate coincides with the probability of necessity from the literature on probabilities of causation. We identify both persuasion rates under a no-backlash condition and a parallel-trends assumption imposed on a known transformation of response probabilities, taking the identity link as the baseline and nonlinear links as sensitivity checks. We develop estimation and inference using GMM and a limited-information method. We demonstrate the usefulness of our framework with an application to a Chinese curriculum reform introduced across provinces at different times.



\raggedbottom

\clearpage

\section{Introduction}\label{section:intro}
\renewcommand{\thefootnote}{\arabic{footnote}\hspace{0.1ex}}
\setcounter{footnote}{0}

Since the dawn of civilization, the meaning of persuasion has evolved from Aristotle's \textit{Rhetoric} to Jane Austen's \textit{Persuasion} and to curriculum reform in China \citep[e.g.,][]{cantoni2017curriculum}. Persuasion is integral to both democracy and the market economy. Economists have empirically measured how persuasive efforts influence the actions of consumers, voters, donors, and investors \citep[see][for a survey of the early literature]{dellavigna2010persuasion}.

Persuasive interventions are often evaluated by asking whether exposure changed average behavior.  Did a media outlet change votes?  Did a new curriculum change political attitudes?  These are average treatment-effect questions in a heterogeneous population.  Persuasion also raises conditional causal questions. Among individuals who took the target action after exposure, was the message necessary for that action?  Looking forward, among individuals who would not have taken the target action without exposure, how many were induced by the message?

A widely used measure of persuasive effectiveness is the persuasion rate, introduced by \citet{dellavigna2007fox} in studying biased news media and voter behavior.  This measure has become a standard metric for comparing persuasion
effects across settings; for example, Table~1 in \citet{dellavigna2010persuasion} and Figure~7 in \citet{Bursztyn:Yang:22}
compile persuasion-rate estimates from numerous studies.  More recently, \citet{jun2023identifying} formalize persuasion rates as causal parameters under both exogenous treatment and instrumental-variable frameworks.  Many empirical applications, however, are observational and are analyzed using difference-in-differences (DID) designs.  This paper asks how persuasion rates should be defined, identified, estimated, and aggregated in DID settings.

A natural example is the Chinese high-school curriculum reform studied by \citet{cantoni2017curriculum}.  The reform changed the content of political education and was introduced across provinces at different times.  The data are
a cross-sectional survey with variation by province and high-school entry cohort, rather than an individual panel with fixed effects.  The original study estimates average effects and reports persuasion rates by rescaling DID estimates of
treatment effects for outcomes measuring political attitudes and beliefs.  This application naturally raises a broader question: when treatment timing is staggered and outcomes are binary, what causal object is delivered by this
DID-based rescaling?  Although the curriculum reform is our main motivating application, the question is general and arises whenever persuasion outcomes are binary and treatment timing varies across groups or cohorts. Examples range from changes in newspaper endorsements in a two-period design \citep{Ladd:Lenz:09} to staggered settings involving village-level cable expansion \citep{jensen2009power}, exposure to small-family role models in television dramas \citep{laferrara2012soap}, and acquisitions of local television stations \citep{levendusky2022sinclair}.

The curriculum example makes clear why average effects can be incomplete measures of persuasion. An average effect on the treated represents the change in the share of treated students who give a reform-aligned response. But this change can be small either because few students could have been persuaded by the curriculum (the size of the potentially persuadable group) or because the curriculum persuaded only a small share of those students (its effectiveness within that group). The two cases can have different policy implications: better targeting may be called for in the former, whereas modifying or replacing the intervention may be more appropriate in the latter. To distinguish these cases, one needs to specify the population relative to which persuasion is measured. If the new curriculum increases the share of students giving a reform-aligned response, one can ask at least two different persuasion questions.  The forward question asks: among students who would not have given that response under the old curriculum, what fraction are induced to give it by exposure to the new curriculum?  The backward question starts from exposed students who give the reform-aligned response and asks what fraction would not have given it under the old curriculum. The second question is Pearl's probability of necessity \citep{pearl1999probabilities}: among observed target responses after exposure, which ones were caused by the treatment?  In contrast, the forward question is not Pearl's probability of sufficiency, which would condition on untreated individuals who could be moved by treatment and is not generally identified by DID comparisons.

These questions are closely related to average treatment effects on the treated (ATT), but they are not the same objects.  Under a no-backlash condition, ATT is the share of treated units moved by the message.  The forward rate divides this causal numerator by the share of treated units who would not have given the target response without exposure, thereby removing cases of ``preaching to the converted.''  The backward rate divides the same numerator by the share of exposed students who give the target response, thereby asking which observed responses were caused by exposure and excluding cases in which the message was effectively ``talking to a brick wall.''  Thus, average effects measure how many people are moved, while persuasion rates measure either the effectiveness of the message among those open to persuasion or the causal attribution of observed target responses. When the no-backlash condition is relaxed and the relevant ATT is nonnegative, the ATT-based ratios are the sharp Fr\'{e}chet--Hoeffding lower bounds on the corresponding persuasion rates. If the ATT is negative, the sharp lower bound is zero.

The framework below develops forward and backward persuasion rates for DID designs with binary outcomes and staggered adoption of a binary treatment. It defines cohort-time rates and their event-study aggregations. Under no backlash,
the underlying cell probabilities also recover the shares of already-persuaded, treatment-persuadable, and never-persuadable types within each treated cohort-time cell, giving the event-study persuasion rates a direct population interpretation.

The central identification problem is to recover untreated counterfactual response probabilities for treated cohort-time cells.  Because the outcome is binary, this requires specifying the scale on which untreated cell probabilities
follow parallel trends.  We impose parallel-trends restrictions on a known transformation of cell probabilities, not on an individual nonlinear panel model with fixed effects.\footnote{For identification in short-panel fixed-effects
binary-choice models and for partial identification of average causal effects in fixed-effects logit models, see \citet{davezies2023fixed} and \citet{davezies2025identification}, respectively. Those models specify individual binary responses through a latent-index structure, whereas our approach imposes parallel trends directly on a known transformation of group-time response probabilities.} Under no backlash, recovering these counterfactual probabilities directly identifies both persuasion rates.

The scale on which untreated probabilities are assumed to evolve is a substantive modeling choice; see \citet{roth2023parallel} on the functional-form sensitivity of parallel trends. Our identification and estimation results hold for any known, strictly increasing link.  In the empirical analysis we take the identity link as the baseline, because it imposes parallel trends directly on response probabilities and is the transparent counterpart of the linear difference-in-differences specification underlying \citet{cantoni2017curriculum}.  We treat alternative scales, in particular the logit link, as a sensitivity analysis: we characterize how much the implied persuasion rates can differ across scales and report identity- and logit-link estimates side by side, so that the role of the scale assumption is explicit rather than hidden. In the simplest design of a single treated cohort observed over two periods, the rescaling used in \citet{cantoni2017curriculum} has the same ratio form as the identity-link forward persuasion rate, but its pooled fitted denominator need not equal the treated cohort's counterfactual denominator.

The paper brings causal persuasion-rate parameters into DID and staggered-adoption settings. To our knowledge, prior work has not developed forward and backward persuasion rates for staggered DID designs. \Cref{sec:background-applications} relates this contribution to the literature on probabilities of causation, modern DID, and adjacent research designs.

The known-link part of the identification argument is also related to nonlinear DID approaches such as \citet{wooldridge2023simple}, who studies ATT identification under transformed parallel trends.  Our target parameters
differ: the ATT is their common numerator, but the forward and backward rates normalize that numerator by conditioning on different subpopulations; only the forward denominator depends on an untreated counterfactual probability.

The empirical analysis returns to the Chinese curriculum reform.  We estimate cohort-time and event-study persuasion rates for political-attitude outcomes and compare them with average treatment effects and with the persuasion calculations
reported in the original study. The goal is constructive: the application shows how a standard staggered DID design can be used to study not only an average treatment effect but also forward and backward persuasion rates.

The rest of the paper proceeds as follows. \Cref{sec:background-applications} presents the motivating applications and
situates the contribution within the persuasion and DID literatures. \Cref{sec:setup-parameters,sec:identification-cell} define forward and backward persuasion rates in staggered designs, introduce event-study aggregations, and give identification results under known-link parallel trends and no backlash. \Cref{sec:est:inf,section:limited:inference} develop estimation and inference, including a GMM estimator that jointly estimates the cell means and cohort shares, with
delta-method inference and a limited-information method based on a reported ATT confidence interval and the share of treated units who do not take the action of interest.  \Cref{section:example2} returns to the empirical application.
\Cref{section:conclusion} summarizes the main findings. Proofs of the main text results are collected in \Cref{app:proofs}. \Cref{app:general-link-covariates} gives the covariate-adjusted version of the estimator, covering panel data, repeated cross-sections, and single-survey cohort designs under a known link function.

\section{Motivating Applications and Related Literature}
\label{sec:background-applications}

\subsection{Curriculum Reform and Ideology}
\label{sec:motiv}

\citet{cantoni2017curriculum} study a Chinese high-school curriculum reform that changed the content of political education and was introduced across provinces at different times.  Their survey is cross-sectional, but it covers Peking
University undergraduates from four high-school entry cohorts, 2006--2009. Because exposure to the new curriculum is determined by province and entry cohort, the empirical variation can be summarized by province-by-entry-cohort cells.

To focus on the persuasion rate, consider a binary outcome $Y_{it}$ equal to one when student $i$ gives the response consistent with the direction of the new curriculum.  We refer to $Y_{it}=1$ as the reform-aligned response.  For this binary outcome, their baseline specification can be written as \citep[][equation (1), p.~361]{cantoni2017curriculum}:
\begin{align}\label{eq:ccyyz}
Y_{it}
= \sum_{t'} \gamma_{t'}\,\one\{t = t' \}
+ \sum_{p} \delta_{p}\,\one\{P_i = p\}
+ \beta\,\text{New Curriculum}_{it}
+ \varepsilon_{it},
\end{align}
where $i$ indexes individuals, $t$ denotes the high-school entry cohort, $P_i$ is the province of high-school attendance, and $\text{New Curriculum}_{it}$ is an indicator equal to one if student $i$ was exposed to the reformed curriculum. The terms $\gamma_t$ and $\delta_p$ are cohort and province fixed effects, and $\beta$ captures the effect of exposure to the new curriculum on the probability of giving the reform-aligned response.

\citet{cantoni2017curriculum} report ``persuasion rates'' by rescaling the estimated effect of the new curriculum \citep[Table~3, column~(7), and p.~382]{cantoni2017curriculum}. Each reported persuasion rate has the form of a ratio: the numerator is $\widehat{\beta}$, and the denominator is constructed from fitted probabilities under the old curriculum. Specifically, after predicting $Y_{it}$ from the regression above, they subtract $\widehat{\beta}$ from the fitted value for students exposed to the new curriculum, leave the fitted value unchanged for students under the old curriculum, and average the resulting fitted counterfactual probabilities.  They then use one minus this average fitted probability as the denominator.  In short, a reported persuasion rate is the estimated coefficient $\widehat{\beta}$ divided by the average fitted probability of not giving the reform-aligned response under the old curriculum.

In fact, \citet{cantoni2017curriculum} anticipate conceptual difficulty with this denominator. They note that ``a more correct definition of the persuasion rate would require us to divide the effect of the new curriculum by the share of students without the desired attitude among individuals who studied under the old curriculum,'' but caution that ``the compositional differences (by province and cohort) in the sample between old and new curriculum students would distort the results'' \citep[p.~382]{cantoni2017curriculum}. Constructing explicit untreated counterfactual probabilities for the relevant treatment-timing cells, as we do below, is one way to address this compositional concern.

The calculation in \citet{cantoni2017curriculum} poses a forward question: among students who would not have given the reform-aligned response under the old curriculum, what fraction are induced to give it by exposure to the new curriculum?  Motivated by the same application, we also study a backward question: among exposed students who give the reform-aligned response, what fraction would not have given it under the old curriculum?

The persuasion-rate calculation in \citet{cantoni2017curriculum} motivates our analysis, but their work is not designed to be a formal identification analysis for a persuasion-rate estimand in a staggered DID design.  Their discussion of a ``conditional'' persuasion rate \citep[p.~382]{cantoni2017curriculum} motivates a formal causal interpretation of such a calculation, especially in a staggered design with binary outcomes. The remainder of this section places the curriculum example alongside other DID applications and relates our contribution to the persuasion and DID literatures.

\subsection{Other Closely Aligned DID Applications}
\label{sec:related-applications}

Political-media persuasion offers a particularly transparent two-period application. \citet{Ladd:Lenz:09} exploit British newspapers' abrupt shifts toward Labour before the 1997 election and compare changes in voting for Labour among readers of switching newspapers with changes among similar respondents who did not read those newspapers. This application fits the two-group, two-period version of our framework: readers of the switching newspapers form the treated group, respondents who did not read them form the comparison group, and outcomes are observed once before and once after the endorsement shifts.

Staggered adoption extends the research design beyond the two-group, two-period case and encompasses a broader range of substantive applications. \citet{jensen2009power} study the staggered introduction of cable television across rural Indian villages, including binary outcomes measuring gender attitudes and household behavior. \citet{laferrara2012soap} exploit the staggered expansion of Rede Globo across Brazilian markets and study whether exposure to small-family role models in television dramas affects the binary decision to give birth in a given year. \citet{levendusky2022sinclair} uses the timing of Sinclair's acquisitions of local television stations to study presidential approval and vote choice in repeated cross-sections. Each application therefore has the group-time structure of our framework: the relevant binary outcomes define cell probabilities, villages or geographic areas define groups, and staggered rollout dates determine treatment timing.

These studies estimate treatment effects rather than the forward and backward persuasion rates developed here. In such designs, no anticipation and known-link parallel trends recover the untreated counterfactual probabilities from observed binary cell probabilities and the treatment-timing distribution. Under no backlash, the two persuasion rates are then identified.

\subsection{Persuasion Rates and the Probability of Necessity}
\label{sec:persuasion-literature}

The forward persuasion rate originates in empirical work measuring the effectiveness of persuasive messages among individuals who would not otherwise take the target action \citep{dellavigna2007fox,dellavigna2010persuasion}. Recent work formulates this rate as a causal parameter and studies its identification under exogenous treatment and instrumental-variable assumptions \citep{jun2023identifying}.

The backward persuasion rate coincides with Pearl's probability of necessity \citep{pearl1999probabilities}; related formulations appear in \citet{Yamamoto:2012,Dawid2014fitting,cinelli2021generalizing,Dawid2022effects,Ding:2024:arXiv:prob_necessity}. As causal parameters, both rates depend on the joint distribution of treated and untreated potential responses \citep{heckman1997making,mullahy2018individual} and can be expressed through shares of latent response types. Related work studies persuasion types, sample selection, and persuasion-like functionals in broader treatment-effect models \citep{yu2023binary,possebom2022probability,kaji2023assessing,ji2023model}.

Two companion papers study persuasion effects in televised-debate and regression-discontinuity designs \citep{jun2025tv,jun2025rdd}.\footnote{These companion papers were developed after the first version of the present study was posted on arXiv at \url{https://arxiv.org/abs/2410.14871}, extending the analysis of persuasion effects to other research designs.} The present paper instead develops cohort-time and event-study persuasion rates for DID, shows how both rates depend on a common causal numerator but different denominator populations, and provides identification and inference under known-link parallel trends.

\subsection{Staggered DID and Event-Study Parameters}
\label{sec:did-literature}

The DID component builds on work identifying treatment effects with panel or repeated-cross-section data \citep{Abadie:05,Ding_Li_2019,arkhangelsky2021synthetic}. Modern staggered-DID methods define cohort-time treatment effects and make comparison-group and aggregation choices explicit \citep{CS:2021,athey2022design}. Related work shows how treatment-effect heterogeneity complicates the interpretation of conventional two-way fixed-effects and event-study regressions \citep{de2020two,goodman2021difference,sun2021estimating}. \citet{NBERw29170} provide a focused treatment of visualization, identifying assumptions, and estimation in linear panel event-study designs. Complementary work studies pre-event trends, heterogeneous coefficients, and variation in exposure in panel event-study designs \citep{AER:event-study-design,Sun:Shapiro:22}.

Our aggregation problem is related but distinct. A cohort-time persuasion rate is a ratio rather than an average treatment effect. Consequently, an event-study persuasion rate can be formed either by aggregating the causal numerators and the relevant denominator populations across eligible cohorts before taking their ratio or, equivalently, as an appropriately weighted average of cohort-time persuasion rates. Because forward and backward rates condition on different populations, they use different aggregation weights.

For recent surveys and practitioner-oriented treatments of modern DID, see \citet{CD:23:survey,roth2023s,baker2026did}. \citet{dechaisemartin2026book} provide a comprehensive textbook treatment. These references cover canonical and staggered designs, comparison-group and aggregation choices, inference, and broader treatment paths; they do not develop the persuasion-rate estimands considered here.

\subsection{Applications Beyond Persuasion}
\label{sec:applications-beyond-persuasion}

The framework also applies beyond settings naturally described as persuasion. \citet{Gentzkow:06} exploits staggered television entry to study voter turnout, while \citet{agrawal2008bitnet} study how staggered Bitnet adoption affects research collaboration. In both settings, the relevant response can be represented as a binary action under a binary absorbing treatment. The forward rate measures the share of units induced to take an action that they would not otherwise take, while the backward rate measures the share of observed actions attributable to treatment.

\Cref{sec:setup-parameters} then defines forward and backward persuasion rates at the group-time level and develops their event-study aggregations.

\section{The Setup and Parameters}\label{sec:setup-parameters}

\subsection{Forward and Backward Persuasion Rates on the Treated}
Let $t \in \{0, 1, 2, \ldots, T\}$ index time and $D_{it}$ indicate whether individual $i$ is exposed to the persuasive treatment at time $t$.   Initially ($t = 0$), no unit is treated. If treatment starts for unit $i$ at time $s \in \{1, 2, \ldots, T\}$, then $D_{it} = 0$ for $t < s$ and $D_{it} = 1$ for $t \geq s$. Never-treated units satisfy $D_{it} = 0$ for all $t$.

Let $\mathscr{S}_i := \bigl\{t\in\{1,\ldots,T\}:\ D_{it} = 1 \bigr\}$, and define $S_i$ by $S_i := \min \mathscr{S}_i$ if $\mathscr{S}_i$ is non-empty, and $S_i:=\infty$ otherwise. Thus, $S_i \in \{1, 2, \ldots, T, \infty\}$ denotes the period in which treatment begins, with $S_i = \infty$ indicating that the unit is never treated. To allow for heterogeneous treatment effects, define $Y_{it}(s)$ as the potential outcome at time $t$ under treatment starting in period $s$, and define $Y_{it}(\infty)$ as the potential outcome at time $t$ if treatment never occurs. For expositional clarity, we first present the case without covariates. Covariate-adjusted estimation and inference are developed in \Cref{sec:est:cov} and \Cref{app:general-link-covariates}. We start with the following assumption.

\begin{assumption}[Staggered Adoption with No Anticipation]\label{ass:stagger-did} At $t = 0$, no unit receives or anticipates treatment, so $Y_{i0} := Y_{i0}(\infty)$ is observed. Further, no one anticipates future treatment: for all $s \in \{1,\dots,T\}$ and $t < s$, $Y_{it}(s) = Y_{it}(\infty)$ almost surely. For $t = 1, \dots, T$, the observed outcome is $Y_{it} := Y_{it}(S_i)$. Moreover, for each treated cohort $s$ and post-treatment period $t\geq s$,
\[
\min \Bigl\{ \Pr\bigl( Y_{it}(\infty) = 0,\ S_i = s \bigr),\
\Pr\bigl( Y_{it} = 1,\ S_i = s \bigr) \Bigr\}
>
0.
\]
\end{assumption}

The positive-probability requirement ensures that the causal parameters below are well-defined. Although the notation is individual-level, with the index $i$ used to define population probabilities and the potential-outcome notation defining the causal parameters, the empirical framework does not require an individual panel. The primitive observables are cell shares (means) of responses together with the distribution of treatment timing, not individual transition histories. These cell shares may come from repeated cross-sections over calendar time, or from a single cross-sectional survey with cohort variation, as in the curriculum example, where province-by-entry-cohort cells play the role of group-by-time cells. The identifying restrictions and estimators in the paper will be stated in terms of counterfactual cell probabilities. This distinction matters because parallel-trends restrictions will be imposed on aggregate probabilities using an assumed link function; it is not an assumption about a nonlinear panel model with individual fixed effects.

In the curriculum example, no anticipation is supported by how the reform was introduced. New textbooks were assigned to incoming senior-high-school (grades 10--12 in China) cohorts, so a student's curriculum was fixed by province and entry year, and students entering one year apart could face entirely different curricula \citep[p.~344]{cantoni2017curriculum}. Older, pre-reform cohorts were not partially treated, because the college entrance examination was based on either
the old or the new curriculum rather than a mixture \citep[p.~344]{cantoni2017curriculum}. Consistent with treatment being determined at the province-by-cohort level, about 94 percent of surveyed students identified as their high-school textbook the one predicted by their province and entry cohort \citep[p.~379]{cantoni2017curriculum}.

For a treated cohort $s$ and post-treatment period $t\geq s$, the forward persuasion rate on the treated (FPRT) is
\[
\theta^{(F)}(s,t)
:=
\Pr\{Y_{it}(s)=1 \mid Y_{it}(\infty)=0,\ S_i=s\}.
\]
It asks: among units in cohort $s$ who would not have given the target response at time $t$ without exposure, what fraction are induced to give the target response at time $t$? In the curriculum example, $\theta^{(F)}(s,t)$ measures the fraction of exposed students in cohort $s$ induced by the new curriculum to give the reform-aligned response among those who would not have given that response under the old curriculum at time $t$. This is the on-the-treated version of the persuasion rate introduced by \citet{dellavigna2007fox} and formalized as a causal parameter by \citet{jun2023identifying}.

The backward persuasion rate on the treated (BPRT) is
\[
\theta^{(B)}(s,t)
:=
\Pr\{Y_{it}(\infty)=0 \mid Y_{it}(s)=1,\ S_i=s\}.
\]
It asks the reverse attribution question: among treated units in cohort $s$ who give the target response at time $t$, what fraction would not have given that response without exposure?  In the curriculum example, $\theta^{(B)}(s,t)$ starts from exposed students who give the reform-aligned response at time $t$ and asks what fraction would not have given that response under the old curriculum.

One drawback of FPRT is that it conditions on a counterfactual subgroup that is not directly observed. BPRT avoids this particular denominator problem: for treated cohort $s$ in period $t\geq s$, the event $\{Y_{it}(s)=1,\ S_i=s\}$ is observed as $\{Y_{it}=1,\ S_i=s\}$.  Its numerator still depends on the counterfactual untreated outcome, which is the object to be identified by the DID restrictions below. In fact, BPRT is the cohort-time analog of the probability of necessity (PN) of \citet{pearl1999probabilities}, useful in legal and policy contexts to assess causation; see also  \citet{Yamamoto:2012} and \citet{Dawid2014fitting}. In contrast, FPRT is not Pearl's probability of sufficiency (PS).  The corresponding PS object in this setting would be
\[
\textrm{PS}(s,t) := \Pr\{ Y_{it}(s) = 1 \ \big|\  Y_{it}(\infty) = 0, S_i = \infty \},
\]
which concerns persuasion among untreated individuals.   The DID parallel-trends assumption identifies counterfactual untreated outcomes for treated units, not counterfactual treated outcomes for untreated units.  Thus, we do not study $\textrm{PS}(s,t)$ in the paper.

\subsection{Event-Study Persuasion Rates}\label{sec:espr}

The FPRT and BPRT introduced in the previous subsection are specific to a treated cohort and period, $(s,t)$. In order to assess dynamic persuasion effects, we consider aggregating across cohorts while fixing the number of periods since treatment. For each event time $j$, let $\mathcal S_j$ denote the set of treated cohorts $s$ such that $s+j$ is observed and $\Pr(S_i=s)>0$. Specifically, for $j=0,1,\ldots,T-1$ with $\mathcal S_j\neq\emptyset$, we define the following parameters:
\begin{align*}
\theta^{(F)}_{\ES}(j) &:= \Pr\{ Y_{i, S_i+j}(S_i) = 1\mid Y_{i,S_i+j}(\infty) = 0, S_i\in\mathcal S_j \},\\
\theta^{(B)}_{\ES}(j) &:= \Pr\{ Y_{i,S_i+j}(\infty) = 0\mid Y_{i,S_i+j}(S_i) = 1, S_i\in\mathcal S_j \},
\end{align*}
which we refer to as the forward and backward event-study persuasion rates (FES and BES), respectively.

The event-study persuasion rates can be obtained by aggregating the cohort-time persuasion rates, i.e., FPRT and BPRT. First, we focus on FES:
\begin{align*}
\theta^{(F)}_{\ES}(j)
&=
\frac{\sum_{s\in\mathcal S_j} \Pr\{ Y_{i, s+j}(s) = 1, Y_{i,s+j}(\infty) = 0, S_i = s \}}{\sum_{s\in\mathcal S_j} \Pr\{ Y_{i,s+j}(\infty) = 0, S_i = s \}}
\\
&=
\frac{\sum_{s\in\mathcal S_j} \Pr\{Y_{i,s+j}(\infty) = 0, S_i = s \} \theta^{(F)}(s,s+j)}{\sum_{s\in\mathcal S_j} \Pr\{ Y_{i,s+j}(\infty) = 0, S_i = s \}}.
\end{align*}
That is, $\theta^{(F)}_{\ES}(j)$ is a weighted average of $\theta^{(F)}(s,s+j)$ across $s$. The BES parameter has an analogous weighted-average representation:
\begin{align*}
\theta^{(B)}_{\ES}(j)
=
\frac{\sum_{s\in\mathcal S_j} \Pr\{Y_{i,s+j}=1,\ S_i=s\}\theta^{(B)}(s,s+j)}
{\sum_{s\in\mathcal S_j} \Pr\{Y_{i,s+j}=1,\ S_i=s\}}.
\end{align*}
In the BES representation, the event-study weights are directly identified from observable treated-cell response shares. In contrast, the FES weights depend on counterfactual untreated response probabilities and are identified only after the DID step below.

\subsection{Persuasion Types among the Treated}
In this subsection, we relate the persuasion rates defined above with the underlying types of the treated population.
Define the shares of four latent response types among the treated population: already-persuaded (AP), never-persuadable (NP), treatment-persuadable (TP), and treatment-contrarian (TC). For $t\geq s$,
\begin{align*}
\AP(s,t) &:= \Pr\{ Y_{it}(s)= 1, Y_{it}(\infty) = 1 \mid S_i = s\},\\
\NP(s,t) &:= \Pr\{ Y_{it}(s)= 0, Y_{it}(\infty) = 0 \mid S_i = s\}, \\
\TP(s,t) &:= \Pr\{ Y_{it}(s)= 1, Y_{it}(\infty) = 0 \mid S_i = s\}, \\
\TC(s,t) &:= \Pr\{ Y_{it}(s)= 0, Y_{it}(\infty) = 1 \mid S_i = s\}.
\end{align*}
This classification parallels the latent type framework in the local average treatment effect (LATE) literature. If we
reinterpret the binary outcome as a ``treatment'' and the treatment indicator as an ``instrument,'' NP, AP, TP, and TC correspond to never-takers, always-takers, compliers, and defiers respectively. Under this analogy, persuasion measures the extent to which treatment causally induces compliance with the targeted behavior. Specifically,
\[
\theta^{(F)}(s,t) = \frac{\TP(s,t)}{\TP(s,t) + \NP(s,t)}
\quad \text{and} \quad
\theta^{(B)}(s,t) = \frac{\TP(s,t)}{\TP(s,t) + \AP(s,t)}.
\]
Therefore, FPRT is the share of TP among NP and TP, whereas BPRT is the share of TP among AP and TP.

In terms of identification, the key difficulty is the common numerator $\TP(s,t)$. The denominators of FPRT and BPRT are marginal probabilities, whereas $\TP(s,t)$ depends on the joint distribution of $Y_{it}(s)$ and $Y_{it}(\infty)$ given $S_i=s$, and these two potential outcomes are never observed together. The following assumption is helpful.
\begin{assumption}[No Backlash]\label{ass:mtr}
For $t\geq s$, we have $\Pr\{ Y_{it}(s) \geq Y_{it}(\infty) \mid S_i = s\} = 1$.
\end{assumption}

\Cref{ass:mtr} is a monotonicity restriction in the target-response direction: exposure may leave a treated unit's response unchanged or move it toward the target response, but it cannot move the unit away from the target response. See, for example, \citet{pearl1999probabilities} and \citet{jun2023identifying}. Since \Cref{ass:mtr} is equivalent to $\TC(s,t) = 0$, the treated population then consists of the AP, NP, and TP types. The restriction also leads to
\begin{multline}\label{eq:tp-and-att}
\TP(s,t)
=
\TP(s,t) - \TC(s,t)
\\
=
\Pr\{ Y_{it}(s) = 1\mid S_i=s \} - \Pr\{ Y_{it}(\infty) = 1\mid S_i=s \} =: \att(s,t),
\end{multline}
where $\att(s,t)$ is the cohort-time average treatment effect among units first treated in period $s$. Therefore,
\begin{align}\label{eq:f,b,att}
\theta^{(F)}(s,t)
&= \frac{\att(s,t)}{\Pr\{Y_{it}(\infty)=0\mid S_i = s\}} \notag\\
&= \frac{\att(s,t)}{\att(s,t) + \Pr\{ Y_{it}(s) = 0\mid S_i = s\}}
= \frac{\att(s,t)}{\att(s,t) + \Pr\{ Y_{it} = 0\mid S_i = s\}},
\notag\\
\theta^{(B)}(s,t)
&= \frac{\att(s,t)}{\Pr\{ Y_{it}(s) = 1 \mid S_i = s\}}
= \frac{\att(s,t)}{\Pr\{ Y_{it} = 1 \mid S_i = s\}}.
\end{align}

\Cref{ass:mtr} can be restrictive in some situations. It rules out the possibility that exposure to the persuasive message causes a behavioral shift in the opposite direction due to distrust or annoyance. For example, if the message is overly manipulative, then it may cause resistance rather than persuasion, thereby violating \Cref{ass:mtr}.

If \Cref{ass:mtr} fails, then $\att(s,t)=\TP(s,t)-\TC(s,t)$, so the cohort-time ATT is weakly below $\TP(s,t)$. Hence, when $\att(s,t)\geq 0$, the right-hand side expressions in \eqref{eq:f,b,att} should be interpreted as lower bounds on the
left-hand side expressions. These lower bounds coincide with the sharp Fr\'{e}chet--Hoeffding lower bounds obtained from the marginals of $Y_{it}(s)$ and $Y_{it}(\infty)$ given $S_i=s$. If $\att(s,t)<0$, the sharp lower bound on $\TP(s,t)$ is zero, so the corresponding lower bound on each persuasion rate is zero rather than the negative ATT-based ratio. See, for example, \citet{jun2023identifying}.

In the curriculum example, no backlash requires that exposure to the new curriculum never move a student away from the reform-aligned response. The estimated effects reported by \citet{cantoni2017curriculum} are in the government-intended direction or statistically insignificant, with no significant reversals, which is consistent with little net backlash on average. However, no backlash can nonetheless fail at the individual level if some students react against a curriculum they perceive as coercive; in that case, when the ATT is nonnegative, the right-hand sides of \eqref{eq:f,b,att} are the sharp Fr\'{e}chet--Hoeffding lower bounds. If the ATT is negative, the sharp lower bound is zero.

\section{Identification Through Cell Probabilities}\label{sec:identification-cell}

For observed response cells, write
\[
\mu_{s,t}:=\Pr\{Y_{it}=1\mid S_i=s\},
\qquad s\in\{1,\ldots,T,\infty\},\quad t=0,\ldots,T,
\]
and, for treated cohorts, write
\[
\pi_s:=\Pr(S_i=s),\qquad s=1,\ldots,T.
\]
We reserve explicit probability notation for potential-outcome quantities. In
particular, let
\[
\tau(s,t)
:=
\Pr\{Y_{it}(\infty)=1\mid S_i=s\},
\qquad t\geq s,
\]
denote the untreated counterfactual response probability for treated cohort
$s$ in period $t$. It follows from \eqref{eq:f,b,att} that, under no backlash,
the common numerator of FPRT and BPRT is
$
\att(s,t)
=
\mu_{s,t}-\tau(s,t).
$
Therefore, identification of the persuasion rates is reduced to that of
$\tau(s,t)$, which is a cell-level object. To identify it, we first aggregate the binary outcome
within treatment-timing cells to obtain response probabilities, and then impose a
parallel-trends restriction on a chosen scale for those probabilities.

\begin{assumption}[Link-Function Parallel Trends]\label{ass:stagger-common-trend}
Let $I_\Lambda\subseteq [0,1]$, and let
$\Lambda:I_\Lambda\to \Lambda(I_\Lambda)$ be a known strictly increasing link
function.
For every treated cohort $s$ and period $t\geq s$,
\begin{multline*}
\Lambda\bigl[\Pr\{Y_{it}(\infty)=1\mid S_i=s\}\bigr]
-
\Lambda\bigl[\Pr\{Y_{i,s-1}(\infty)=1\mid S_i=s\}\bigr]
\\
=
\Lambda\bigl[\Pr\{Y_{it}(\infty)=1\mid S_i=\infty\}\bigr]
-
\Lambda\bigl[\Pr\{Y_{i,s-1}(\infty)=1\mid S_i=\infty\}\bigr],
\end{multline*}
where the displayed conditional probabilities are well-defined and lie in
$I_\Lambda$.
\end{assumption}

In \Cref{ass:stagger-common-trend}, the control group consists of ``never-treated'' units ($S_i = \infty$), while the treatment group consists of units with $S_i = s$. The assumption compares the potential outcomes of these two groups between the period of interest, $t\geq s$, and the most recent pre-treatment period $s-1$.
This assumption could be strengthened to allow any not-yet-treated group ($S_i = s' > t$) to serve as a control and to use any earlier pre-treatment period ($s-k$ for $k = 1, \ldots, s$) as the comparison period. In this paper,
we adopt the simpler version in \Cref{ass:stagger-common-trend}, focusing on a primary control group and comparison period. For general discussions of alternative parallel trends assumptions, see, for example, \citet{CS:2021,roth2023s}, and for recent advances on efficient estimation under stronger versions of this assumption, see \citet{Chen2025eff}.

In the curriculum example, the credibility of parallel trends rests on design features rather than random assignment: \citet{cantoni2017curriculum} emphasize that adoption timing was not random, since provinces switched only after teachers were trained and materials prepared \citep[p.~345]{cantoni2017curriculum}. The province-by-cohort design absorbs province- and cohort-specific differences \citep[p.~342]{cantoni2017curriculum}, and the event-study evidence shows little sign of differential pre-trends \citep[Fig.~4]{cantoni2017curriculum}, with results robust
to placebo adoption dates \citep[p.~372]{cantoni2017curriculum}. This evidence speaks most directly to the identity-link version of the restriction as \citet{cantoni2017curriculum} use the linear regression model as their main specification.

The choice of $\Lambda$ specifies the scale on which untreated cell
probabilities are assumed to move in parallel.  Our baseline is the identity link denoted by $\Lambda_{\mathrm{id}}$, for which $I_{\Lambda_{\mathrm{id}}}=[0,1]$ and $\Lambda_{\mathrm{id}}(p)=p$. Under the identity link, \Cref{ass:stagger-common-trend} is the usual parallel-trends restriction for response probabilities, and the resulting estimators are the transparent counterpart of linear DID. Nonlinear links serve as the leading alternatives for assessing sensitivity to this scale choice. Two such examples are
(i) $\Lambda_{\log}(p) = \log(p),\, I_{\Lambda_{\log}}=(0,1]$;
(ii) the logit link function denoted by $\Lambda_{\mathrm{logit}}$, for which $I_{\Lambda_{\mathrm{logit}}}=(0,1)$ and
\[
\Lambda_{\mathrm{logit}}(p)=\log\left(\frac{p}{1-p}\right),
\qquad
\Lambda_{\mathrm{logit}}^{-1}(z)=\frac{\exp(z)}{1+\exp(z)}
=\frac{1}{1+\exp(-z)}.
\]
The log link $\Lambda_{\log}$ imposes common proportional growth in untreated target-response probabilities. The logit link $\Lambda_{\mathrm{logit}}$ imposes common proportional growth in untreated odds.

The general known-link formulation is close in spirit to \citet{wooldridge2023simple}, who studies staggered DID under parallel trends for known transformations of untreated conditional means. His target parameter is ATT. In our setting, the same type of restriction identifies the untreated counterfactual probability that enters both the ATT numerator and the FPRT denominator. The logit link also connects to the compositional DID framework of \citet{Boussim:Comp}: in the binary case, $\log\{p/(1-p)\}$ is the log ratio of the target-response share to the non-target-response share, so logit parallel trends impose parallel evolution of the log odds rather than of the raw probabilities.

Given a link function $\Lambda$, define the following population quantity,
which is directly identified from observed cell probabilities:
\begin{equation}\label{eq:psi-lambda}
\psi_{s,t}(\Lambda)
:=
\Lambda^{-1}\left\{
\Lambda(\mu_{s,s-1})
+\Lambda(\mu_{\infty,t})
-\Lambda(\mu_{\infty,s-1})
\right\}.
\end{equation}

\begin{theorem}[Identification of Staggered Persuasion Rates]\label{thm:id-cell}
Under \Cref{ass:stagger-did,ass:stagger-common-trend}, for every treated cohort $s$ and
period $t\geq s$, $\tau(s,t)$ is identified by $\psi_{s,t}(\Lambda)$. If \Cref{ass:mtr} holds in addition, then the FPRT and BPRT under the maintained link $\Lambda$ are
\begin{align*}
\theta^{(F)}(s,t;\Lambda)
=
\frac{\mu_{s,t}-\psi_{s,t}(\Lambda)}
{1-\psi_{s,t}(\Lambda)},\quad
\theta^{(B)}(s,t;\Lambda)
=
\frac{\mu_{s,t}-\psi_{s,t}(\Lambda)}
{\mu_{s,t}}.
\end{align*}
For event time $j$ with $\mathcal S_j\neq\emptyset$,
\begin{align*}
\theta^{(F)}_{\ES}(j;\Lambda)
=
\frac{
\sum_{s\in\mathcal S_j}\pi_s
\{\mu_{s,s+j}-\psi_{s,s+j}(\Lambda)\}
}{
\sum_{s\in\mathcal S_j}\pi_s
\{1-\psi_{s,s+j}(\Lambda)\}
},\quad
\theta^{(B)}_{\ES}(j;\Lambda)
=
\frac{
\sum_{s\in\mathcal S_j}\pi_s
\{\mu_{s,s+j}-\psi_{s,s+j}(\Lambda)\}
}{
\sum_{s\in\mathcal S_j}\pi_s\mu_{s,s+j}
}.
\end{align*}
\end{theorem}

The argument $\Lambda$ refers to the maintained scale on which untreated cell probabilities are assumed to follow parallel trends. With the correct link, the quantities in \Cref{thm:id-cell} coincide with the causal parameters defined in \Cref{sec:espr}. The theorem is stated in terms of observed cell probabilities and cohort shares. The link function is
applied after aggregation within cohort-by-period cells indexed by $(s,t)$. Consequently, the same identification logic can be used with panel data, repeated cross-sections, or a single cross-sectional survey with cohort variation, provided the relevant cell probabilities are observed or consistently estimated. Because persuasion rates are conditional probabilities, the event-study formulas first aggregate the relevant numerators and denominators over $\mathcal S_j$ by the law of total probability and then take their ratio.\footnote{This is reminiscent of, for example, \citet{frolich2007nonparametric}; aggregating the local average treatment effect across compliers amounts to averaging the numerator and denominator of the conditional Wald estimand.}

\begin{corollary}[Identification of Persuasion Types]\label{cor:types} Under \Cref{ass:stagger-did,ass:mtr,ass:stagger-common-trend}, for every treated cohort $s$ and period $t\geq s$, the type shares among cohort $s$ at time $t$ are point-identified by
\[
\AP(s,t)=\psi_{s,t}(\Lambda),
\quad
\NP(s,t)=1-\mu_{s,t},
\quad
\TP(s,t)=\mu_{s,t}-\psi_{s,t}(\Lambda),
\quad
\TC(s,t)=0.
\]
\end{corollary}

\Cref{cor:types} re-expresses \Cref{thm:id-cell} in terms of the latent types: no backlash sets $\TC=0$ and equates the treatment-persuadable share $\TP$ with the cohort-time ATT, monotonicity makes the already-persuaded share equal to the untreated counterfactual response probability $\AP(s,t)=\tau(s,t)$, and the never-persuadable share equals the treated non-response share $\NP(s,t)=1-\mu_{s,t}$.

These shares give the event-study parameters of \Cref{sec:espr} a transparent population meaning. Aggregating the type shares over eligible cohorts,
\begin{align}\label{eq:share:event}
\begin{split}
\AP_{\ES}(j)&=\frac{\sum_{s\in\mathcal S_j}\pi_s\psi_{s,s+j}(\Lambda)}
{\sum_{s\in\mathcal S_j}\pi_s},
\quad
\NP_{\ES}(j)=\frac{\sum_{s\in\mathcal S_j}\pi_s(1-\mu_{s,s+j})}
{\sum_{s\in\mathcal S_j}\pi_s}, \\
\TP_{\ES}(j)&=\frac{\sum_{s\in\mathcal S_j}\pi_s\{\mu_{s,s+j}-\psi_{s,s+j}(\Lambda)\}}
{\sum_{s\in\mathcal S_j}\pi_s},
\end{split}
\end{align}
yields the type composition of the pooled population of eligible treated cohorts observed $j$ periods after treatment. The event-study rates are exactly the type fractions of that pooled population,
\[
\theta^{(F)}_{\ES}(j;\Lambda)=\frac{\TP_{\ES}(j)}{\TP_{\ES}(j)+\NP_{\ES}(j)}
\quad \text{ and } \quad
\theta^{(B)}_{\ES}(j;\Lambda)=\frac{\TP_{\ES}(j)}{\TP_{\ES}(j)+\AP_{\ES}(j)}.
\]
This representation extends directly to other aggregation schemes that reweight the eligible cohorts.

\subsection{\texorpdfstring{Relation to \citet{cantoni2017curriculum}}{Relation to Cantoni et al. (2017)}}

The identity-link estimand specializes cleanly in the simplest design, where it connects directly to the persuasion rate reported by \citet{cantoni2017curriculum}.

\begin{corollary}[Two-Period Identity-Link Benchmark]\label{cor:2x2-cantoni}
Consider a single treated cohort $s$ with the never-treated group as control, observed in the two periods $s-1$ and $s$. Under \Cref{ass:stagger-did,ass:mtr,ass:stagger-common-trend} with the identity link,
\[
\theta^{(F)}(s,s;\Lambda_{\mathrm{id}})
=
\frac{\att(s,s)}{1-\psi_{s,s}(\Lambda_{\mathrm{id}})},
\qquad
\theta^{(B)}(s,s;\Lambda_{\mathrm{id}})
=
\frac{\att(s,s)}{\mu_{s,s}},
\]
where
\[
\att(s,s)
=
(\mu_{s,s}-\mu_{s,s-1})-(\mu_{\infty,s}-\mu_{\infty,s-1})
\]
is the two-period difference-in-differences contrast and $\psi_{s,s}(\Lambda_{\mathrm{id}})=\mu_{s,s-1}+\mu_{\infty,s}-\mu_{\infty,s-1}$.
\end{corollary}

\Cref{cor:2x2-cantoni} is the identity-link, single-cohort, two-period specialization of \Cref{thm:id-cell}: the numerator is the standard difference-in-differences contrast, and $1-\psi_{s,s}(\Lambda_{\mathrm{id}})$ is the share of the treated cohort that would not give the reform-aligned response absent treatment.

\Cref{cor:2x2-cantoni} makes precise the sense in which the forward persuasion rate generalizes the calculation in \citet{cantoni2017curriculum}. Let $\widehat Y_{it}$ denote the fitted value from the linear probability model in \eqref{eq:ccyyz}. \citet{cantoni2017curriculum} form
$
\widetilde Y_{it}
:=\widehat Y_{it} -\widehat\beta\,\textrm{New Curriculum}_{it}
$
and report $\widehat\beta/\{1-\overline{\widetilde Y}\}$, where
$\overline{\widetilde Y}$ is the sample mean of $\widetilde{Y}_{it}$. In a saturated two-group, two-period design with $S_i\in\{1,\infty\}$ indexing the treated and comparison groups and $t\in\{0,1\}$, the four fitted cells satisfy
$
\widehat\mu_{\infty,t}=\widehat\gamma_t+\widehat\delta_\infty
\text{ and }
\widehat\mu_{1,t}=\widehat\gamma_t+\widehat\delta_1
+\widehat\beta\one\{t=1\}.
$
Thus, $\widehat\beta$ is the usual DID estimator for $\att(1,1)$ and $\widehat\psi_{1,1}(\Lambda_{\mathrm{id}})=\widehat\mu_{1,1}-\widehat\beta$, so the FPRT estimator is $\widehat\beta/(1-\widehat\mu_{1,1}+\widehat\beta)$. The calculation in \citet{cantoni2017curriculum} coincides with FPRT only if $\overline{\widetilde Y}=\widehat\mu_{1,1}-\widehat\beta$, which is not implied by the design because $\widetilde Y_{it}$ equals that quantity only in the treated post-treatment cell.

More generally, two features of the staggered binary setting prevent a single
rescaled coefficient from recovering the corresponding event-study rate.
First, with several adoption cohorts the two-way fixed-effects coefficient is a weighted combination of cohort-time effects that need not equal any average treatment effect \citep[e.g.,][]{de2020two,CS:2021,goodman2021difference}, whereas the numerator here is the cohort-time $\att(s,t)$. Second, the denominator must use the cohort-time counterfactual share $1-\psi_{s,t}(\Lambda)$, and we must aggregate numerators and denominators separately across cohorts, rather than dividing by one pooled fitted share that mixes cohort and province composition.

\subsection{Identity- and Logit-Link Counterfactuals}

The identification result above allows the untreated response probability to be constructed under any chosen link. In applications, the main empirical comparison is often between the identity link, which imposes parallel trends on the probability scale, and the logit link, which imposes parallel trends on the log-odds scale. Both specifications use the same observed cell probabilities and differ only in the constructed untreated counterfactual. Their comparison is therefore a sensitivity analysis for cell-level counterfactual construction, not a comparison between individual nonlinear panel estimators.

For a treated cohort $s$ and event time $j$ with $t=s+j$, define the identity-link and logit-link versions of the counterfactual untreated response probability by
\begin{align*}
\psi_{s,s+j}(\Lambda_{\mathrm{id}})
&:=
\mu_{s,s-1}+\mu_{\infty,s+j}-\mu_{\infty,s-1},\\
\Lambda_{\mathrm{logit}}\{\psi_{s,s+j}(\Lambda_{\mathrm{logit}})\}
&:=
\Lambda_{\mathrm{logit}}(\mu_{s,s-1})
+
\Lambda_{\mathrm{logit}}(\mu_{\infty,s+j})
-
\Lambda_{\mathrm{logit}}(\mu_{\infty,s-1}).
\end{align*}
Define
\begin{align*}
\Delta_\psi(s,s+j)
&:=
\psi_{s,s+j}(\Lambda_{\mathrm{logit}})-\psi_{s,s+j}(\Lambda_{\mathrm{id}}),
\end{align*}
which is the gap between the two link-based untreated counterfactual probabilities.

If identity-link parallel trends hold for the untreated response probabilities but the logit link is used instead, $\Delta_\psi(s,s+j)$ is the bias in the constructed untreated counterfactual. The sign is reversed when logit-link parallel trends hold and the identity link is used instead.
For each $\ell\in\{\mathrm{id},\mathrm{logit}\}$ and $r\in\{F,B\}$, define $\theta^{(r)}_{\ell}(s,s+j):=\theta^{(r)}(s,s+j;\Lambda_\ell)$ whenever the relevant denominator is nonzero.
To state the comparison compactly, define
\begin{align*}
N_\psi(s,s+j)
&:=
\{\mu_{s,s-1}-\mu_{\infty,s-1}\}
\{\mu_{\infty,s+j}-\mu_{\infty,s-1}\}
\{1-\mu_{s,s-1}-\mu_{\infty,s+j}\},\\
D_\psi(s,s+j)
&:=
\mu_{s,s-1}\mu_{\infty,s+j}(1-\mu_{\infty,s-1})
+
(1-\mu_{s,s-1})(1-\mu_{\infty,s+j})\mu_{\infty,s-1}.
\end{align*}

\begin{theorem}[Persuasion-Rate Differences]
\label{thm:id-logit-rate-gap}
Suppose that $\mu_{s,s-1}$, $\mu_{\infty,s+j}$, and $\mu_{\infty,s-1}$ lie in
$(0,1)$,
$\mu_{s,s+j}\in(0,1)$,
$1-\psi_{s,s+j}(\Lambda_{\mathrm{id}})>0$,
and
$1-\psi_{s,s+j}(\Lambda_{\mathrm{logit}})>0$.
Then
\[
\Delta_\psi(s,s+j)
=
\frac{N_\psi(s,s+j)}{D_\psi(s,s+j)}.
\]
Also,
\begin{align*}
\theta^{(F)}_{\mathrm{logit}}(s,s+j)-\theta^{(F)}_{\mathrm{id}}(s,s+j)
&=
-\frac{\Delta_\psi(s,s+j)(1-\mu_{s,s+j})}
{\{1-\psi_{s,s+j}(\Lambda_{\mathrm{id}})\}\{1-\psi_{s,s+j}(\Lambda_{\mathrm{logit}})\}}, \\
\theta^{(B)}_{\mathrm{logit}}(s,s+j)-\theta^{(B)}_{\mathrm{id}}(s,s+j)
&=
-\frac{\Delta_\psi(s,s+j)}{\mu_{s,s+j}}.
\end{align*}
\end{theorem}

\Cref{thm:id-logit-rate-gap} shows how the misspecification error in untreated response probabilities propagates into FPRT and BPRT. For either persuasion rate, the logit-minus-identity gap is the bias from using the logit link when identity-link parallel trends hold; conversely, the identity-minus-logit gap is the bias from using the identity link when logit-link parallel trends hold. The FPRT gap reverses the sign of the counterfactual gap and rescales its magnitude by the ratio of the treated-cell nonresponse share to the product of the two link-specific forward denominators; it can therefore be magnified when either denominator term is small. The BPRT gap likewise reverses the sign and rescales its magnitude by the inverse of the observed treated-cell response probability. In \Cref{thm:id-logit-rate-gap}, the signs of the logit-minus-identity gaps for FPRT and BPRT are the opposite of $\operatorname{sign}\{N_\psi(s,s+j)\}$: for $r\in \{F,B\}$,
\[
\operatorname{sign}
\{\theta^{(r)}_{\mathrm{logit}}(s,s+j)-\theta^{(r)}_{\mathrm{id}}(s,s+j)\}
=
-\operatorname{sign}\{N_\psi(s,s+j)\}.
\]
The three factors in $N_\psi(s,s+j)$ have direct interpretations: baseline treated-control imbalance ($\mu_{s,s-1}-\mu_{\infty,s-1}$), the control-group trend ($\mu_{\infty,s+j}-\mu_{\infty,s-1}$), and a binary-outcome boundary term
($1-\mu_{s,s-1}-\mu_{\infty,s+j}$). If $N_\psi(s,s+j)>0$, the logit counterfactual is larger than the identity counterfactual, so the logit-link FPRT and BPRT are smaller than their identity-link counterparts. If $N_\psi(s,s+j)<0$, the logit counterfactual is smaller and the logit-link persuasion rates are larger. If $N_\psi(s,s+j)$ is close to zero and $D_\psi(s,s+j)$ is bounded away from zero, the two counterfactuals are close. In particular, the two counterfactuals coincide when there is no baseline imbalance, no control-group trend, or $\mu_{s,s-1}+\mu_{\infty,s+j}=1$.

There is also an admissibility issue. The identity-link counterfactual $\psi_{s,s+j}(\Lambda_{\mathrm{id}})$ can fall outside $[0,1]$, in which case it is not an admissible probability for that cell. The logit-link counterfactual remains in $(0,1)$ whenever the observed cell probabilities entering the formula are interior. This admissibility issue is one reason to report the logit-link estimates alongside the identity-link baseline: the identity link is the transparent counterpart of linear DID but can produce inadmissible counterfactuals, whereas the logit counterfactual is always interior.

Reporting identity-link and logit-link estimates side by side separates sampling uncertainty, which is conditional on a chosen link, from sensitivity to the scale on which untreated response probabilities are assumed to follow parallel trends. If the two estimates are close relative to their sampling uncertainty, the substantive conclusion is robust to this scale choice. If they differ materially, the data alone do not determine the persuasion rate without specifying the scale on which untreated response probabilities follow parallel trends.

\section{Estimation and Inference via GMM}\label{sec:est:inf}

Our main objects of interest are the forward and backward event-study persuasion rates constructed under a chosen known link function $\Lambda$:
\[
\theta^{(F)}_{\ES}(j;\Lambda)
\quad\text{and}\quad
\theta^{(B)}_{\ES}(j;\Lambda),
\]
where we explicitly show dependence on $\Lambda$.  Throughout, we maintain the identification conditions in \cref{ass:stagger-did,ass:mtr,ass:stagger-common-trend}.

To estimate the observed cell probabilities $\mu_{s,t}$ used in \Cref{sec:identification-cell}, we use notation that covers panel data, repeated cross-sections, and single-survey cohort designs. For $s\in\{1,\ldots,T,\infty\}$ and period $t\in\{0,\ldots,T\}$, let $C_{i,st}$ be the indicator that observation $i$ contributes to the cohort-time cell $(s,t)$, and let $Y_{i,st}^{\mathrm{obs}}$ denote the observed binary response used for that cell. In this notation, $\mu_{s,t}$ is the conditional cell mean $\Pr(Y_{i,st}^{\mathrm{obs}}=1\mid C_{i,st}=1)$ and is identified by
\[
\Exp\left[C_{i,st}\{Y_{i,st}^{\mathrm{obs}}-\mu_{s,t}\}\right]=0.
\]
Equivalently, the normalized sample analog is
\[
\widehat\mu_{s,t}
=
\frac{\sum_{i=1}^n C_{i,st}Y_{i,st}^{\mathrm{obs}}}
{\sum_{i=1}^n C_{i,st}},
\]
whenever the denominator is nonzero. Thus, the displayed moment is just the cell-mean condition written before dividing by the cell probability. In panel notation, $C_{i,st}=\one\{S_i=s\}$ and $Y_{i,st}^{\mathrm{obs}}=Y_{it}$; in repeated-cross-section or single-survey cell notation, $C_{i,st}=\one\{S_i=s,T_i=t\}$ and $Y_{i,st}^{\mathrm{obs}}=Y_i$, where $T_i$ indexes the observation's calendar or cohort time.

The cohort shares $\pi_s$ are estimated along with the cell means. For estimation, let $\widehat{\mathcal S}_j$ denote the sample eligible set: cohorts in $\mathcal S_j$ for which the cells needed for $(s,s+j)$, $(s,s-1)$, $(\infty,s+j)$, and $(\infty,s-1)$ are nonempty and the corresponding estimated cell means lie in the domain of the chosen link. We also require the constructed $\widehat\psi_{s,s+j}(\Lambda)$ to lie in $[0,1]$ and the cohort-time denominators $1-\widehat\psi_{s,s+j}(\Lambda)$ and $\widehat\mu_{s,s+j}$ to be strictly positive. The population formulas for the cohort-time and event-study persuasion rates under the maintained link $\Lambda$ are given in \Cref{thm:id-cell}. The estimation problem is therefore to estimate the cell means and cohort shares entering $\psi_{s,s+j}(\Lambda)$ and the event-study aggregations.

For the identity link, we have
\[
\psi_{s,s+j}(\Lambda_{\mathrm{id}})
=
\mu_{s,s-1}+\mu_{\infty,s+j}-\mu_{\infty,s-1},
\]
so $\mu_{s,s+j}-\psi_{s,s+j}(\Lambda_{\mathrm{id}})$ is the usual DID contrast for the cohort-time ATT, and $1-\psi_{s,s+j}(\Lambda_{\mathrm{id}})$ is the denominator of the FPRT. Hence, the identity-link estimator preserves the direct bridge to linear DID and to the rescaling used in the motivating example. We accordingly report identity-link estimates as the baseline and logit-link estimates as the leading sensitivity check.

For the logit link, provided that the three cell probabilities entering \eqref{eq:psi-lambda} are interior, we have
\[
\psi_{s,s+j}(\Lambda_{\mathrm{logit}})
=
\frac{
\mu_{s,s-1}\mu_{\infty,s+j}(1-\mu_{\infty,s-1})
}{
\mu_{s,s-1}\mu_{\infty,s+j}(1-\mu_{\infty,s-1})
+
(1-\mu_{s,s-1})(1-\mu_{\infty,s+j})\mu_{\infty,s-1}
}.
\]
This counterfactual lies in $(0,1)$ whenever the observed cell probabilities in the formula lie in $(0,1)$.

Let $\vartheta$ collect the cell means $\{\mu_{s,t}\}$ for all nonempty cells used in the analysis and the treated-cohort shares $\{\pi_s\}_{s=1}^T$. The data vector contains the observed response, cell-membership indicators, and treatment-timing information. The just-identified moment vector stacks
\[
\bigl\{
C_{i,st}\{Y_{i,st}^{\mathrm{obs}}-\mu_{s,t}\}
\bigr\}_{(s,t)\ \mathrm{nonempty}}
\ \cup\
\bigl\{\one\{S_i=s\}-\pi_s\bigr\}_{s=1}^T.
\]
The GMM estimator $\widehat{\vartheta}$ solves the corresponding sample moment conditions; $\widehat{\mu}_{s,t}$ is the sample mean in cell $(s,t)$, and $\widehat{\pi}_s$ is the sample share of cohort $s$. For a chosen link $\Lambda$, define $\widehat{\psi}_{s,s+j}(\Lambda)$ by evaluating \eqref{eq:psi-lambda} at $t=s+j$ after replacing each population cell mean with its sample analog. The plug-in estimators are
\begin{align*}
\widehat{\theta}^{(F)}_{\ES}(j;\Lambda)
=
\frac{
\sum_{s\in\widehat{\mathcal S}_j}\widehat\pi_s
\{\widehat\mu_{s,s+j}-\widehat\psi_{s,s+j}(\Lambda)\}
}{
\sum_{s\in\widehat{\mathcal S}_j}\widehat\pi_s
\{1-\widehat\psi_{s,s+j}(\Lambda)\}
},\quad
\widehat{\theta}^{(B)}_{\ES}(j;\Lambda)
=
\frac{
\sum_{s\in\widehat{\mathcal S}_j}\widehat\pi_s
\{\widehat\mu_{s,s+j}-\widehat\psi_{s,s+j}(\Lambda)\}
}{
\sum_{s\in\widehat{\mathcal S}_j}\widehat\pi_s\widehat\mu_{s,s+j}
}.
\end{align*}

Under standard regularity conditions for just-identified GMM with clustering at the assignment level, the joint estimator $\widehat\vartheta$ is asymptotically normal. Assume also that $\widehat{\mathcal S}_j=\mathcal S_j$ with probability approaching one. This stability condition holds, for example, when, uniformly over $s\in\mathcal S_j$, the population cell means lie in the interior of the link domain, the constructed counterfactuals are bounded away from the boundary of $[0,1]$, and the relevant cohort-time denominators are bounded away from zero. If, in addition, $\Lambda$ and $\Lambda^{-1}$ are continuously differentiable in neighborhoods of the relevant population values and the event-study denominators are bounded away from zero, the displayed plug-in estimators are smooth functions of $\widehat\vartheta$ and are therefore
asymptotically normal by the delta method. Standard errors can be obtained by applying the delta method directly to the displayed ratios, using the cluster-robust covariance matrix for $\widehat\vartheta$. We cluster at the level of treatment assignment, such as province in \Cref{section:example2}.

The persuasion-type shares are estimated on the same basis. By \Cref{cor:types}, the cohort-time shares are $\widehat\AP(s,t)=\widehat\psi_{s,t}(\Lambda)$, $\widehat\NP(s,t)=1-\widehat\mu_{s,t}$, and $\widehat\TP(s,t)=\widehat\mu_{s,t}-\widehat\psi_{s,t}(\Lambda)$, and the event-study aggregates replace the population quantities in \eqref{eq:share:event} with the corresponding sample analogs. Each of these is a smooth function of the joint estimator $\widehat\vartheta$, so it is asymptotically normal by the delta method.

\begin{remark}[Alternative Inference]
Persuasion-rate estimators are ratios and can be unstable when their denominators are close to zero. A null hypothesis
$H_0:\theta^{(F)}_{\ES}(j;\Lambda)=\theta_{F,H}$, for a given value $\theta_{F,H}$, can be tested through the equivalent smooth restriction
\[
\sum_{s\in\mathcal S_j}
\pi_s
\left[
\mu_{s,s+j}-\psi_{s,s+j}(\Lambda)
-\theta_{F,H}\{1-\psi_{s,s+j}(\Lambda)\}
\right]=0.
\]
Similarly, $H_0:\theta^{(B)}_{\ES}(j;\Lambda)=\theta_{B,H}$, for a given value $\theta_{B,H}$, can be tested through
\[
\sum_{s\in\mathcal S_j}
\pi_s
\left[
\mu_{s,s+j}-\psi_{s,s+j}(\Lambda)
-\theta_{B,H}\mu_{s,s+j}
\right]=0,
\]
which avoids relying directly on the delta method for the ratio. In sample implementation, the population quantities in these restrictions are replaced by their plug-in estimates, and $\mathcal S_j$ is replaced by $\widehat{\mathcal S}_j$. Anderson-Rubin-type confidence sets can be constructed by inverting these pointwise tests.
\end{remark}

\begin{remark}[Pre-Treatment Parallel-Trends Diagnostics]\label{rmk:pretrend-diagnostic}
For pre-treatment diagnostics, it is preferable to examine link-function parallel trends rather than pre-treatment persuasion-rate ratios. For pre-treatment periods $t<s-1$, the natural diagnostic relative to the baseline period $s-1$ is the link-scale contrast
\[
\{ \Lambda(\mu_{s,t})-\Lambda(\mu_{s,s-1}) \}
-
\{\Lambda(\mu_{\infty,t})-\Lambda(\mu_{\infty,s-1})\},
\]
or an event-time aggregation of such contrasts. Under the identity link, this reduces to the familiar raw-probability placebo DID contrast. When pre-treatment ratio analogs are displayed, they are descriptive placebo summaries and do not have the causal persuasion-rate interpretation of FPRT, BPRT, FES, or BES.
\end{remark}

\subsection{Covariate Adjustment}\label{sec:est:cov}

The GMM estimator above is saturated in cohort-by-period cells and imposes no functional-form restrictions on the corresponding response probabilities. With discrete covariates, one can condition on cells defined by cohort, period, and covariate values and then suitably aggregate over the covariate distribution. With continuously distributed
covariates, however, full saturation is generally infeasible; moreover, additive regression adjustment using two-way fixed effects in a linear or logit model need not recover a convex average of heterogeneous treatment effects \citep[e.g.,][]{NBERw29709,NBERw30108}. \Cref{app:general-link-covariates} therefore develops a covariate-adjusted version of the estimator. It estimates the required conditional cell means given pre-treatment covariates, constructs the counterfactual quantities conditional on those covariates, and averages the resulting numerators and denominators over the covariate distribution accordingly. Because the estimand is written in terms of cell probabilities, the same formulation applies to panel data, repeated cross-sections, and single-survey cohort designs under a known link function.

\section{Limited-Information Inference}\label{section:limited:inference}

This section gives a version of limited-information inference when the researcher does not have enough information to estimate all cell probabilities entering $\psi_{s,t}(\Lambda)$, but does have an estimate of the relevant ATT numerator, together with some information about the treated-cell response share. To keep notation light, this section suppresses dependence on the assumed link function: write $\psi_{s,t}$ for $\psi_{s,t}(\Lambda)$, and interpret all ATT numerators and persuasion rates as constructed under that link.

Consider first a treated cohort $s$ and post-treatment period $t\geq s$. Under \Cref{ass:mtr}, the common numerator of FPRT and BPRT is
\[
\att(s,t)
:=
\mu_{s,t}-\psi_{s,t},
\]
where $\psi_{s,t}$ is the constructed untreated counterfactual. Let
\[
q(s,t):=\Pr(Y_{it}=0\mid S_i=s)=1-\mu_{s,t}
\]
denote the observed share of treated units that do not give the target response in cell $(s,t)$. Using \eqref{eq:f,b,att}, the cohort-time persuasion rates can be written as
\begin{align*}
\theta^{(F)}(s,t)
=
\frac{\att(s,t)}
{\att(s,t)+q(s,t)}
\quad \text{and}\quad
\theta^{(B)}(s,t)
=
\frac{\att(s,t)}
{1-q(s,t)},
\end{align*}
provided $0\leq q(s,t) < 1$ and $\att(s,t)+q(s,t)>0$. These identities are useful when a study reports an estimate of $\att(s,t)$ but does not report enough information to reconstruct the full persuasion-rate estimators. In that case, a confidence interval for $q(s,t)$ can be combined with a confidence interval for $\att(s,t)$.

The same logic applies to event-study persuasion rates, but the target cohorts and weights must be those used by the ATT supplied to the limited-information procedure rather than the full-information eligibility set in \Cref{sec:est:inf}. For each event time $j$, let $\mathcal S^{\mathrm{LI}}_j$ denote the treated cohorts entering that ATT and let $\omega_{s,j}\geq0$, $s\in\mathcal S^{\mathrm{LI}}_j$, be its aggregation weights, normalized so that $\sum_{s\in\mathcal S^{\mathrm{LI}}_j}\omega_{s,j}=1$. Define
\[
\att_{\ES,\mathrm{LI}}(j)
:=
\sum_{s\in\mathcal S^{\mathrm{LI}}_j}\omega_{s,j}
\{\mu_{s,s+j}-\psi_{s,s+j}\}.
\]
Define the corresponding observed treated-cell zero-response share by
\[
q_{\ES,\mathrm{LI}}(j)
:=
\sum_{s\in\mathcal S^{\mathrm{LI}}_j}\omega_{s,j}
(1-\mu_{s,s+j}).
\]
The associated limited-information target rates are
\begin{align*}
\theta^{(F)}_{\ES,\mathrm{LI}}(j)
=
\frac{\att_{\ES,\mathrm{LI}}(j)}
{\att_{\ES,\mathrm{LI}}(j)+q_{\ES,\mathrm{LI}}(j)}
\quad \text{and}\quad
\theta^{(B)}_{\ES,\mathrm{LI}}(j)
=
\frac{\att_{\ES,\mathrm{LI}}(j)}
{1-q_{\ES,\mathrm{LI}}(j)}.
\end{align*}
The event-study rates given in \Cref{sec:est:inf} are the special case with $\mathcal S^{\mathrm{LI}}_j=\mathcal S_j$ and
$\omega_{s,j}=\pi_s/\sum_{\ell\in\mathcal S_j}\pi_\ell$. Other external event-study ATT estimators may imply different cohorts or weights; the corresponding zero-response share must use exactly that same target. For notational economy, we suppress the subscript $\mathrm{LI}$ below and write $\att_{\ES}(j)$, $q_{\ES}(j)$, and $\theta^{(r)}_{\ES}(j)$, $r\in\{F,B\}$, for the specified limited-information target.

Thus limited-information inference requires an estimate or confidence interval for the weighted ATT numerator, such as one obtained from recent staggered-DID methods \citep[e.g.,][]{CS:2021,Chen2025eff}, together with matching treated response information. Let $\widehat\omega_{s,j}$ denote the cohort weights used by the supplied ATT estimate, normalized so that $\sum_{s\in\mathcal S^{\mathrm{LI}}_j}\widehat\omega_{s,j}=1$. When the treated target cell $(s,s+j)$ is nonempty for every cohort receiving positive weight, the natural plug-in estimator is
\[
\widehat q_{\ES}(j)
:=
\sum_{s\in\mathcal S^{\mathrm{LI}}_j}\widehat\omega_{s,j}
(1-\widehat\mu_{s,s+j}).
\]
Unlike the full-information set $\widehat{\mathcal S}_j$ in \Cref{sec:est:inf}, this construction does not require baseline or control cells, or an admissible counterfactual $\widehat\psi_{s,s+j}$. It requires only consistent estimation of the treated-cell means and weights for the cohorts defining the supplied ATT. The same construction applies in panel data, repeated cross-sections, and single-survey cohort designs.

Suppose we are given an estimate $\widehat{\att}_{\ES}(j)$ with standard error $\widehat{\mathrm{se}}_{\att,\ES}(j)$, together with an interval $[\underline q_{\ES}(j),\overline q_{\ES}(j)]\subseteq [0,1]$ for $q_{\ES}(j)$; a normal-approximation confidence interval for $q_{\ES}(j)$ may not be contained in $[0,1]$, but in that case, we intersect it with $[0,1]$ because $q_{\ES}(j)$ is a probability. Fix nominal noncoverage probabilities $\alpha_{\att}\in(0,1)$ and $\alpha_q\in[0,1)$ with $\alpha:=\alpha_{\att}+\alpha_q<1$.  Let $z_\tau$ be the $\tau$-quantile of the standard normal distribution, and consider the following formulas:
\begin{align*}
\widehat L^F_{\ES}(j)
&:=
\frac{\widehat{\att}_{\ES}(j)}
{\widehat{\att}_{\ES}(j)+\overline q_{\ES}(j)}
-
z_{1-\alpha_{\att}/2} \widehat{\mathrm{se}}_{\att,\ES}(j)
\frac{\overline q_{\ES}(j)}
{\{\widehat{\att}_{\ES}(j)+\overline q_{\ES}(j)\}^2},\\
\widehat U^F_{\ES}(j)
&:=
\frac{\widehat{\att}_{\ES}(j)}
{\widehat{\att}_{\ES}(j)+\underline q_{\ES}(j)}
+
z_{1-\alpha_{\att}/2}\widehat{\mathrm{se}}_{\att,\ES}(j)
\frac{\underline q_{\ES}(j)}
{\{\widehat{\att}_{\ES}(j)+\underline q_{\ES}(j)\}^2},\\
\widehat L^B_{\ES}(j)
&:=
\frac{\widehat{\att}_{\ES}(j)}
{1-\underline q_{\ES}(j)}
-
z_{1-\alpha_{\att}/2}\widehat{\mathrm{se}}_{\att,\ES}(j)
\frac{1}{1-\underline q_{\ES}(j)},\\
\widehat U^B_{\ES}(j)
&:=
\frac{\widehat{\att}_{\ES}(j)}
{1-\overline q_{\ES}(j)}
+
z_{1-\alpha_{\att}/2}\widehat{\mathrm{se}}_{\att,\ES}(j)
\frac{1}{1-\overline q_{\ES}(j)}.
\end{align*}
On exceptional events where any of the displayed denominators is zero, define the corresponding confidence limit or standardized statistic arbitrarily. Under the conditions of the theorem we state below, such exceptional events have probability $o(1)$. When $\alpha_q>0$, one admissible choice for
$[\underline q_{\ES}(j),\overline q_{\ES}(j)]$ is a normal-approximation confidence interval based on $\widehat q_{\ES}(j)$ and its standard error. If instead a range for $q_{\ES}(j)$ is imposed as deterministic auxiliary information, one sets $\alpha_q=0$.  \Cref{app:limited-info-details} gives the exact
formula used in the empirical application and discusses finite-sample
zero-denominator events.
The following result formalizes the limited-information confidence intervals
for event-study persuasion rates.

\begin{theorem}[Limited-Information Confidence Intervals for Event-Study Persuasion Rates]
\label{thm:limited-info-ci}
Fix event horizon $j$ and maintain \Cref{ass:stagger-did} for the cohorts and periods used in the limited-information analysis. Let $\alpha_{\att}\in(0,1)$ and $\alpha_q\in[0,1)$ satisfy $\alpha:=\alpha_{\att}+\alpha_q<1$. Suppose that $q_{\ES}(j)<1$ and $\att_{\ES}(j)\geq 0$, as implied by \Cref{ass:stagger-did,ass:mtr}, respectively, and that $\att_{\ES}(j)+q_{\ES}(j)>0$, so that both $\theta^{(F)}_{\ES}(j)$ and $\theta^{(B)}_{\ES}(j)$ are well-defined.  Suppose further that, with probability approaching one, $0\leq\underline q_{\ES}(j)\leq\overline q_{\ES}(j)<1$, and both $\att_{\ES}(j)+\underline q_{\ES}(j)$ and $1-\overline q_{\ES}(j)$ are bounded away from zero.  Assume further that the interval $[\underline q_{\ES}(j), \overline q_{\ES}(j)]$ contains $q_{\ES}(j)$ with probability at least $1-\alpha_q+o(1)$, and that $\widehat{\att}_{\ES}(j)$ is consistent for $\att_{\ES}(j)$, $\widehat{\mathrm{se}}_{\att,\ES}(j)>0$ with probability approaching one, and
\[
\frac{\widehat{\att}_{\ES}(j)-\att_{\ES}(j)}
{\widehat{\mathrm{se}}_{\att,\ES}(j)}
\convd N(0,1).
\]
Then, for $r\in\{F,B\}$,
\[
\Pr\bigl\{\theta^{(r)}_{\ES}(j)\in
[\widehat L^r_{\ES}(j),\widehat U^r_{\ES}(j)] \bigr\}
\geq
1-\alpha + o(1).
\]
\end{theorem}

The cohort-time version is obtained by replacing $\att_{\ES}(j)$ with $\att(s,t)$, $q_{\ES}(j)$ with $q(s,t)$, and $\theta^{(r)}_{\ES}(j)$ with $\theta^{(r)}(s,t)$. \Cref{thm:limited-info-ci} addresses uncertainty about
$\att_{\ES}(j)$ and $q_{\ES}(j)$ by a Bonferroni argument. The parameter $\alpha_q$ is the noncoverage probability allocated to the interval for the treated zero-response share, while $\alpha_{\att}$ is the noncoverage probability allocated to the ATT estimate. Thus, a $(1-\alpha)$ confidence interval is obtained by choosing $\alpha_{\att}$ and $\alpha_q$ so that $\alpha_{\att}+\alpha_q=\alpha$. For example, for a 95 percent interval, one may set $\alpha=.05$ and split the noncoverage probability evenly, or choose a different split if one source of uncertainty is viewed as more important.

The construction resembles confidence intervals for partially identified parameters, such as those in \citet{Imbens/Manski:04} and \citet{Stoye:07}, but the source of the interval is different. In the present setting, the event-study
persuasion rates are point identified from the cell probabilities in \Cref{thm:id-cell}; there is no identified set once the relevant cell information is available. The interval for \(q_{\ES}(j)\) enters only because this section studies a limited-information case in which the researcher does not observe, or cannot reconstruct, all of the cell probabilities. We use Bonferroni because it only requires a coverage statement, or a given sensitivity range, for \(q_{\ES}(j)\), together with the usual inference for the ATT numerator.

\begin{remark}[Back-of-the-Envelope Inference]
A researcher may want to compute the persuasion rates using information available only in a published paper, without having access to the full dataset. Suppose that one has a confidence interval for the ATT numerator, along with auxiliary information on the range of $q_{\ES}(j)$. If the interval for $q_{\ES}(j)$ is viewed as deterministic auxiliary information, one can set $\alpha_q=0$ and allocate the full noncoverage probability to the ATT interval and interpret the resulting interval as inference conditional on the imposed range on $q_{\ES}(j)$. We illustrate both this back-of-the-envelope calculation and an estimated-interval version for the curriculum example in \Cref{sec:example:limited}.
\end{remark}

\section{Empirical Example: Curriculum Persuasion}\label{section:example2}

\citet{cantoni2017curriculum} provide a useful application of our framework. Their Peking University survey is cross-sectional, but it contains variation by province of high-school attendance and high-school entry cohort. We use provinces adopting the new curriculum in 2007, 2008, and 2009 as treated cohorts. Provinces adopting in 2010 serve as the comparison group within the 2006--2009 entry-cohort window, so we code them as $S_i=\infty$ in this empirical implementation. Provinces that adopted before 2007 are excluded.

\begin{table}[htbp]
	\centering
	\begin{threeparttable}
\caption{Staggered Introduction of the New Curriculum\label{tb:ccyyz}}
    \small
	\begin{tabularx}{16cm}{XcX}
	&
\begin{tabular}{lccccccc}
\hline
Group $(S_i)$ by year   & \multicolumn{7}{c}{Event Horizon ($j$)} \\
of a new curriculum & $-4$ & $-3$ & $-2$ & $-1$ & $0$ & $1$ & $2$ \\
\hline
2007        &   &          &          & 2006 & 2007 & 2008 & 2009 \\
2008        &  &          & 2006 & 2007 & 2008 & 2009 &        \\
2009        &   & 2006 & 2007 & 2008 & 2009 &          &       \\
2010 $(S_i = \infty)$   & 2006 & 2007 & 2008 & 2009 &   &  &    \\
\hline
\end{tabular}
	&
	\end{tabularx}
	\vspace{1ex}
	\begin{tablenotes}
	{\footnotesize
	\item[a] The provinces that introduced the new curriculum in 2007 are Beijing, Heilongjiang, Hunan, Jilin, and Shaanxi.
	\item[b] The provinces that introduced the new curriculum in 2008 are Henan, Jiangxi, Shanxi, and Xinjiang.
	\item[c] The provinces that introduced the new curriculum in 2009 are Hebei, Hubei, Inner Mongolia, and Yunnan.
	\item[d] The provinces that introduced the new curriculum in 2010 are Chongqing, Gansu, Guangxi, Guizhou, Qinghai, Sichuan, and Tibet. They define the policy-timing comparison group because the observed 2006--2009 entry cohorts were not exposed to the new curriculum. The survey contains no observations from Tibet, so the realized comparison sample consists of the other six provinces.
	\item[e] An entry in the table refers to the relevant high-school entry year in the cross-sectional survey (i.e., $s+j$) for each $S_i = s$ group.
	\item[f] The provinces that adopted the new curriculum before 2007 are excluded in our analysis.
	}
	\end{tablenotes}
	\end{threeparttable}
\end{table}

We first consider the first outcome variable in \citet[Table~3, first row]{cantoni2017curriculum}, ``Trust: central government.'' For this illustration, the binary outcome equals one for a reform-aligned response to this question. Following their persuasion-rate calculation, we code the response as one when the answer is weakly above the median response for that question. The estimates below use the identity-link implementation, which is closest to the original linear DID rescaling in \citet{cantoni2017curriculum}. Standard errors are clustered at the province level, the level at which curriculum adoption is assigned.

\begin{figure}[!htbp]
\caption{Forward and Backward Persuasion Rates by Year of Treatment Adoption\label{figure-ccyyz-1}}
\vspace*{1ex}
\begin{center}
\makebox{
\includegraphics[scale=0.55]{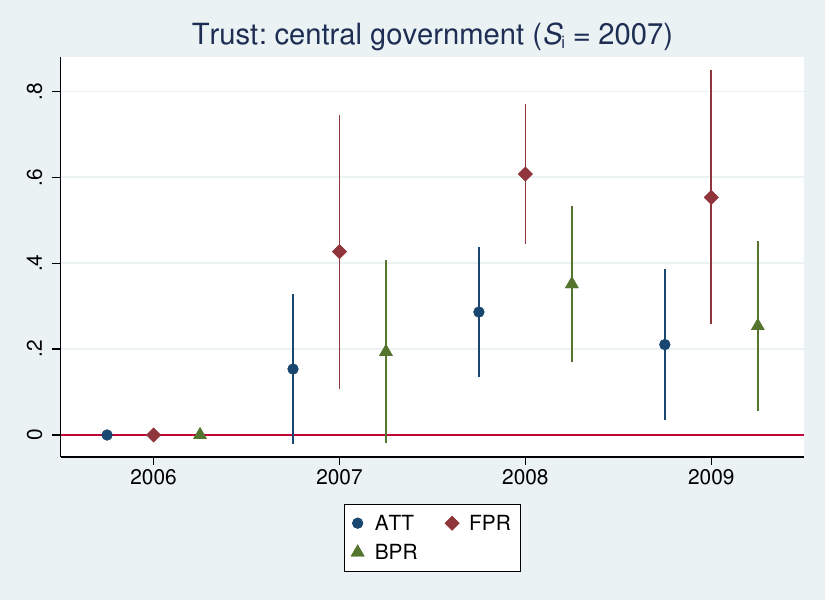}
\includegraphics[scale=0.55]{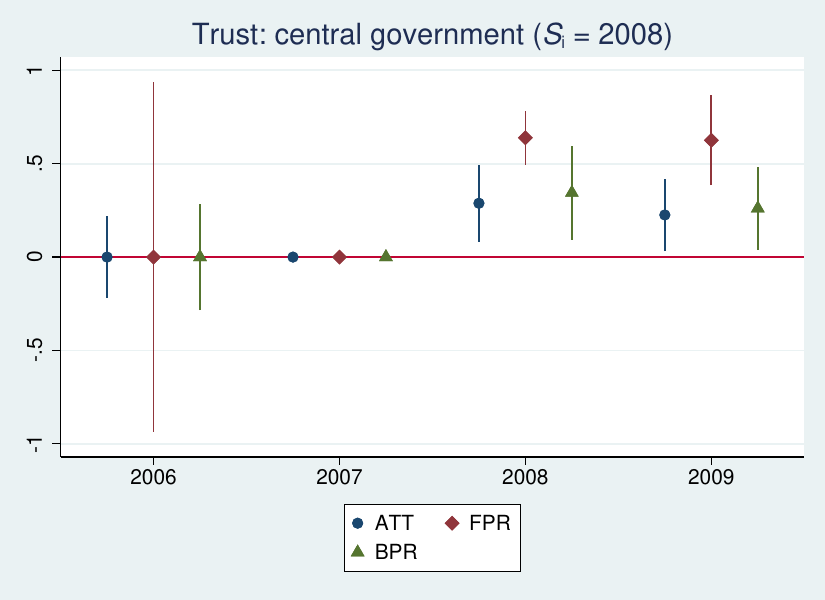}
}
\end{center}
\par
\vspace*{1ex}
\parbox{6in}{\footnotesize
Notes: The left and right panels show estimates for the $S_i = 2007$ and $S_i = 2008$ groups, respectively. At post-treatment horizons, the panels show the average treatment effect on the treated (ATT) and the forward and backward persuasion rates (FPRT and BPRT). Negative-horizon points, when shown, are descriptive placebo ratio analogs rather than causal persuasion-rate estimates. Vertical lines represent 95\% pointwise confidence intervals.}
\end{figure}

\Cref{figure-ccyyz-1} reports cohort-time estimates for the 2007 and 2008 treated cohorts. In the left panel, the $S_i=2007$ group uses 2006 as the pre-treatment baseline and follows the first treated cohort through 2009. In the right panel, the $S_i=2008$ group has two pre-treatment entry cohorts and two post-treatment entry cohorts. For post-treatment points in both panels, FPRT and BPRT rescale the same ATT numerator by different target populations, so they are no smaller than ATT. Each post-treatment cohort-time estimate uses the identity-link formula in \Cref{thm:id-cell}; in the first treated period, this reduces to the two-period benchmark in \Cref{cor:2x2-cantoni}. For example, in the first treated period of the $S_i=2007$ cohort, an ATT of about $0.15$ and an untreated non-response share of about $0.36$ give a forward rate of about $0.43$. In that cell, both ATT and BPRT are statistically insignificant at the 5 percent level, whereas FPRT is significant, underscoring the importance of going beyond ATT. Overall, the cohort-time estimates indicate substantial effects of the textbook reform on students' trust in the central government.

\begin{table}[!htbp]
\centering
\begin{threeparttable}
\caption{Estimated Persuasion-Type Shares: Trust in Central Government\label{tb:type-shares}}
\small
\begin{tabular*}{\textwidth}{@{\extracolsep{\fill}}lccccc@{}}
\hline
& \multicolumn{3}{c}{Cohort $S_i=2007$}
& \multicolumn{2}{c}{Cohort $S_i=2008$} \\
Type / entry year $t$ & 2007 & 2008 & 2009 & 2008 & 2009 \\
\hline
AP & 0.64 (0.08) & 0.53 (0.08) & 0.62 (0.07) & 0.55 (0.11) & 0.64 (0.10) \\
NP & 0.21 (0.03) & 0.18 (0.03) & 0.17 (0.04) & 0.16 (0.02) & 0.14 (0.03) \\
TP & 0.15 (0.09) & 0.29 (0.08) & 0.21 (0.09) & 0.29 (0.10) & 0.23 (0.10) \\
\hline
\end{tabular*}
\begin{tablenotes}
{\footnotesize
\item Notes: Estimated shares of already-persuaded (AP), never-persuadable (NP), and treatment-persuadable (TP) types among each treated cohort for ``Trust: central government,'' under the identity link and no backlash (\Cref{cor:types}). Point estimates sum to one up to rounding; TP equals the cohort-time ATT in \Cref{figure-ccyyz-1}. Cluster-robust standard errors (clustered by province), computed by the delta method, are in parentheses.
}
\end{tablenotes}
\end{threeparttable}
\end{table}

\Cref{tb:type-shares} reports the implied type shares from \Cref{cor:types} for these cohorts. The already-persuaded share is the largest component in every cell, consistent with FPRT exceeding BPRT, while the treatment-persuadable share equals the cohort-time ATT.

\begin{figure}[!htbp]
\caption{Forward and Backward Persuasion Rates (FES and BES)\label{figure-ccyyz-2}}
\vspace*{1ex}
\begin{center}
\makebox{
\includegraphics[scale=0.38]{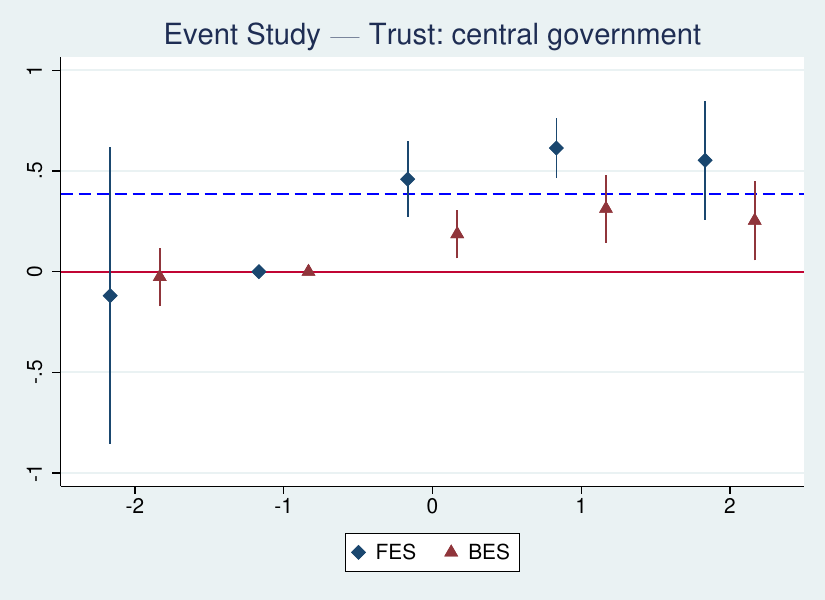}
\includegraphics[scale=0.38]{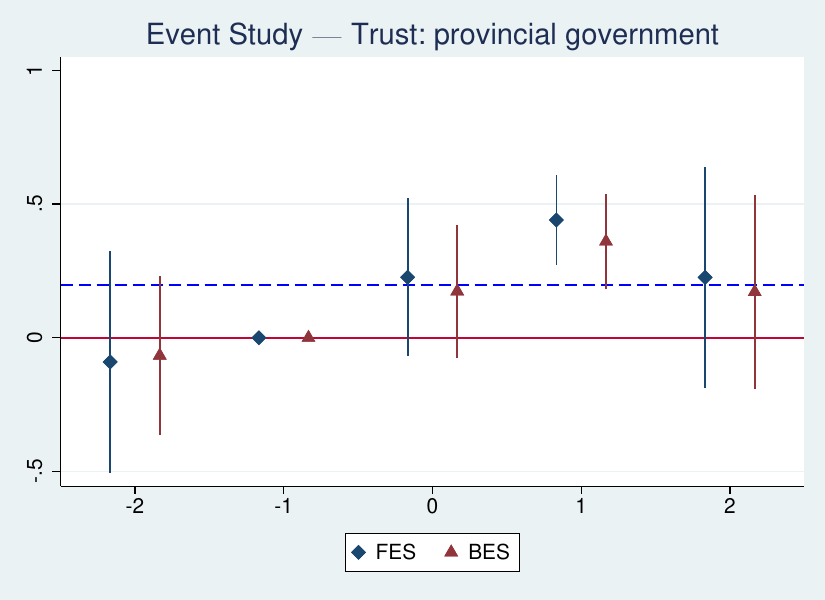}
\includegraphics[scale=0.38]{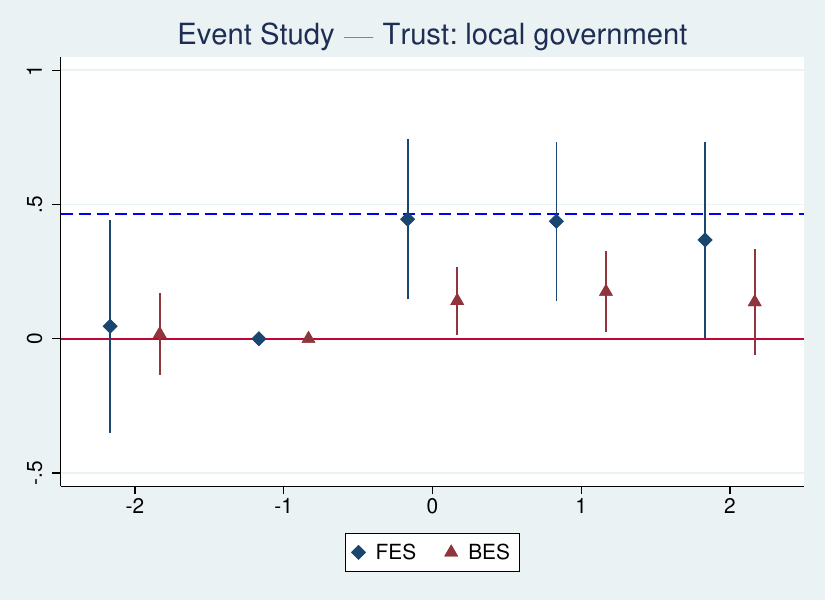}
} \\
\makebox{
\includegraphics[scale=0.38]{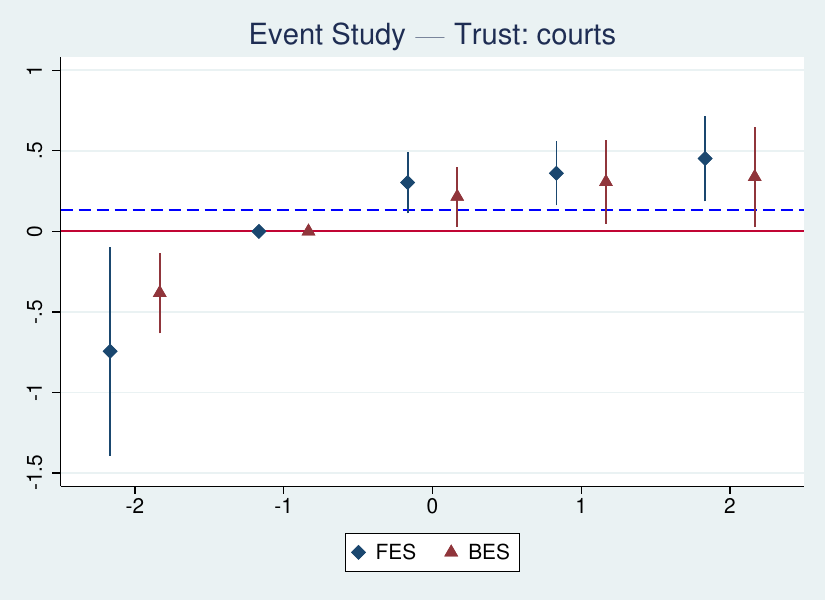}
\includegraphics[scale=0.38]{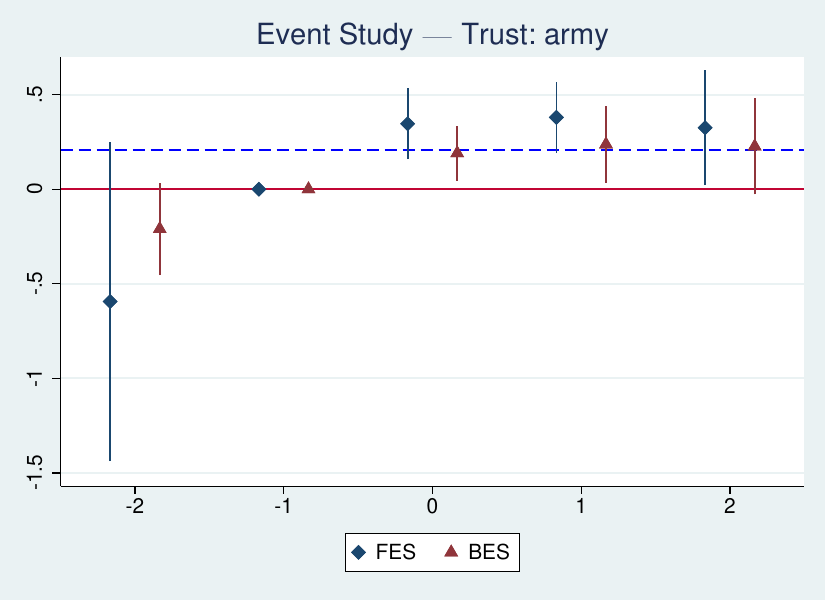}
\includegraphics[scale=0.38]{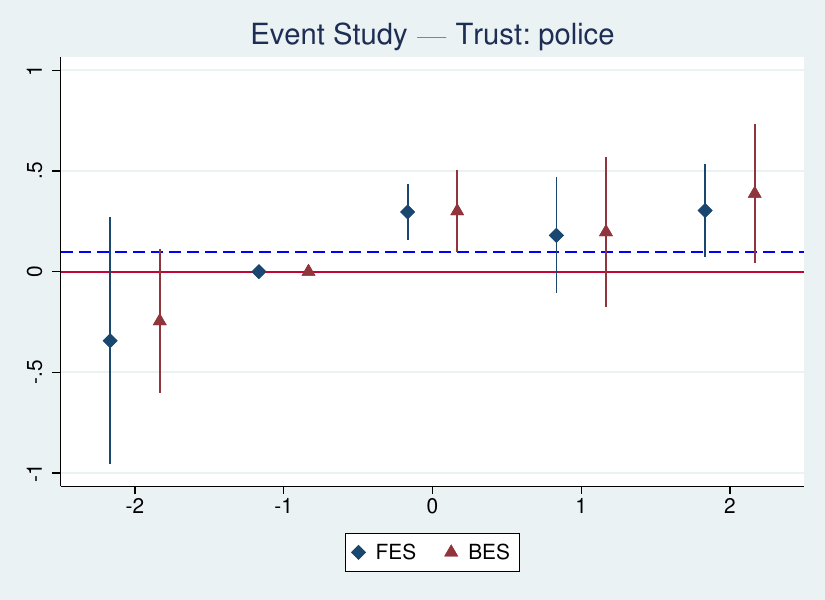}
}
\end{center}
\par
\vspace*{1ex}
\parbox{6in}{\footnotesize
Notes: At post-treatment horizons, each panel shows estimates of the forward and backward event-study persuasion rates (FES and BES). Negative-horizon points are descriptive placebo ratio analogs rather than causal FES or BES estimates. Vertical lines represent 95\% pointwise confidence intervals.}
\end{figure}

We do not emphasize separate estimates for the $S_i=2009$ group because that cohort has only one post-treatment entry cohort and the estimates are imprecise. Instead, \Cref{figure-ccyyz-2} reports post-treatment FES and BES, which aggregate eligible treated cohorts at each event time, together with negative-horizon placebo ratio analogs. The six panels correspond to the first six outcome variables in \citet[Panel~A, Table~3]{cantoni2017curriculum}. Pre-treatment placebo analogs are generally insignificant, except for ``Trust: courts.''\footnote{Pre-treatment ratio analogs can be noisy because the pre-treatment ATT numerator should be close to zero and small denominators can magnify sampling variation. We therefore treat these plotted ratios only as descriptive placebo summaries; the formal diagnostic is the link-scale contrast in \Cref{rmk:pretrend-diagnostic}.} Most post-treatment estimates are positive and pointwise statistically significant.

The blue dashed line in each panel of \Cref{figure-ccyyz-2} is the persuasion rate reported in \citet{cantoni2017curriculum}. Their calculation reports a single rescaled DID coefficient from the regression model in \eqref{eq:ccyyz}, whereas FES aggregates cohort-time numerators and denominators separately before taking the ratio. With heterogeneous treatment effects and with provinces treated before 2007 excluded from the comparison group, the FES estimates are generally above the dashed-line benchmark. BES estimates are typically smaller than FES estimates, consistent with a larger already-persuaded share than never-persuadable share. Overall, the results reaffirm the main finding in \citet{cantoni2017curriculum}: studying the new curriculum led to more positive views of China's governance.

\subsection{Inference from Limited Information}\label{sec:example:limited}

The construction in \Cref{section:limited:inference} provides a confidence
interval for a persuasion rate when only the treatment-effect numerator and
partial information about the treated response share are available, a relevant
case here, since the original study reports point persuasion rates without
confidence intervals \citep[Table 3, Column (7)]{cantoni2017curriculum}. We
illustrate it for the $S_i=2007$ cohort in its first treated period,
using ``Trust: central government.''

\Cref{tb:limited-info} compares the full-information joint-GMM delta method with two calculations based on limited information about $q$. The second row treats a range for $q$ as deterministic auxiliary information; the third combines an estimated interval for $q$ with ATT inference using a Bonferroni split. This is the modular case in which the ATT may instead come from an external staggered-DID estimator while $q$ is estimated from the treated cells. The limited-information intervals are wider because they use coarser information about $q$ and, in the third row, because the Bonferroni split raises the ATT critical value from $1.96$ to $2.24$. Nevertheless, the full- and limited-information intervals are of similar magnitude in this application.

\begin{table}[!htbp]
\centering
\begin{threeparttable}
\caption{Limited-Information Confidence Intervals: Trust in Central Government\label{tb:limited-info}}
\small
\begin{tabular}{lcc}
\hline
Inference & $\theta^{(F)}$ (forward) & $\theta^{(B)}$ (backward) \\
\hline
Full information (delta method) & $[0.11,\ 0.75]$ &
$[0.00,\ 0.41]$ \\
Back-of-the-envelope (assumed $q$ range) & $[0.11,\ 0.79]$ &
$[0.00,\ 0.44]$ \\
ATT CI and estimated $q$ CI (Bonferroni) & $[0.06,\ 0.87]$ &
$[0.00,\ 0.49]$ \\
\hline
\end{tabular}
\begin{tablenotes}
{\footnotesize
\item Notes: 95\% confidence intervals for the forward and backward persuasion
rates of the $S_i=2007$ cohort in its first treated period, ``Trust: central
government,'' under the identity link. The ATT and $q$ estimates are $0.154$ (s.e.\ $0.089$) and $0.206$ (s.e.\ $0.035$), respectively; calculations use unrounded values. Reported intervals are intersected with the parameter space $[0,1]$.  The back-of-the-envelope row imposes $q \in[0.15,0.25]$ as deterministic
($\alpha_q=0$); the last row combines ATT inference with the estimated
$97.5\%$ interval for $q$, $[0.128,0.285]$, using the Bonferroni split $\alpha_{\att}=\alpha_q=0.025$ (\Cref{thm:limited-info-ci}).
}
\end{tablenotes}
\end{threeparttable}
\end{table}

\subsection{Link Sensitivity}

In this application, the identity-link estimates are the baseline empirical implementation because they are closest to the linear DID specification in \citet{cantoni2017curriculum}. The same observed treatment-timing cell probabilities can also be used to construct untreated counterfactuals under the logit link. This comparison changes only the scale on which untreated response probabilities are assumed to follow parallel trends. Reporting both links also addresses a structural limitation of the identity link: its constructed counterfactual probability can fall outside $[0,1]$. In this application, the identity-link counterfactuals are interior for all post-treatment cells used to construct the reported persuasion rates, so the baseline estimates are admissible.

\begin{figure}[htbp]
\caption{Event-Study Persuasion Rates Under Identity and Logit Links\label{figure-ccyyz-link}}
\vspace*{1ex}
\begin{center}
\makebox{
\includegraphics[width=0.48\textwidth]{figures/event_study_central_government.pdf}
\includegraphics[width=0.48\textwidth]{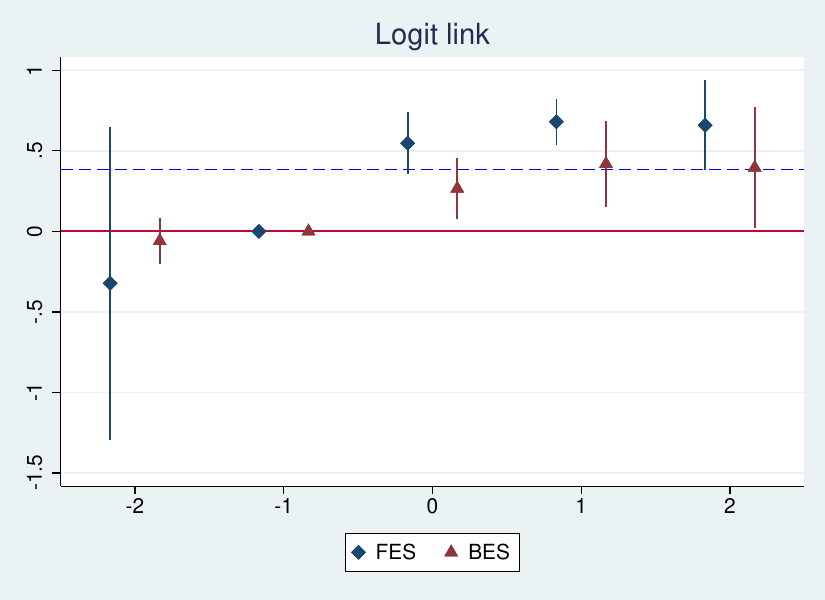}
}
\end{center}
\par
\vspace*{1ex}
\parbox{6in}{\footnotesize
Notes: The left panel repeats the identity-link event-study estimates for ``Trust: central government'' from the first panel of \Cref{figure-ccyyz-2}. The right panel shows the corresponding logit-link estimates. Both panels use the
same observed treatment-timing cell probabilities; only the link used to construct untreated counterfactual probabilities changes. Vertical lines represent 95\% pointwise confidence intervals. The dashed horizontal line is the persuasion rate reported in \citet{cantoni2017curriculum}.}
\end{figure}

\begin{table}[!htbp]
\centering
\begin{threeparttable}
\caption{Decomposition of the Identity-Logit Link Gap\label{tb:link-gap-decomp}}
\small
\begin{tabular}{lrrrrrrrr}
\hline
& Base & Ctrl. & Bound. & & & & & \\
\makecell{Entry\\year} & gap & trend & term & $N_\psi$ & $D_\psi$ & $\Delta_\psi$ & $\Delta_F$ & $\Delta_B$ \\
\hline
2007 & -0.169 & -0.107 & -0.557 & -0.010 & 0.095 & -0.107 & 0.131 & 0.135 \\
2008 & -0.169 & -0.219 & -0.445 & -0.016 & 0.113 & -0.145 & 0.093 & 0.178 \\
2009 & -0.169 & -0.128 & -0.536 & -0.012 & 0.098 & -0.118 & 0.106 & 0.143 \\
\hline
\end{tabular}
\begin{tablenotes}
{\footnotesize
\item Notes: The table reports cohort-time quantities for the $S_i=2007$ adoption group and the outcome ``Trust: central government.'' Base gap is $\mu_{s,s-1}-\mu_{\infty,s-1}$, Ctrl. trend is $\mu_{\infty,s+j}-\mu_{\infty,s-1}$, and Bound. term is $1-\mu_{s,s-1}-\mu_{\infty,s+j}$. Also, $\Delta_\psi=\psi_{s,s+j}(\Lambda_{\mathrm{logit}})-\psi_{s,s+j}(\Lambda_{\mathrm{id}})$, $\Delta_F=\theta^{(F)}_{\mathrm{logit}}(s,s+j)-\theta^{(F)}_{\mathrm{id}}(s,s+j)$, and $\Delta_B=\theta^{(B)}_{\mathrm{logit}}(s,s+j)-\theta^{(B)}_{\mathrm{id}}(s,s+j)$.}
\end{tablenotes}
\end{threeparttable}
\end{table}

\Cref{tb:link-gap-decomp} links the empirical pattern in \Cref{figure-ccyyz-link} to \Cref{thm:id-logit-rate-gap}. In all three post-treatment cells for the 2007 adoption group, the treated baseline share is below the comparison baseline share, the comparison-group response share falls relative to baseline, and the boundary term is negative. Hence, $N_\psi(s,s+j)<0$. Because $D_\psi(s,s+j)>0$, the logit-link counterfactual is below the identity-link counterfactual. This lower counterfactual raises the link-based ATT numerator and makes both FPRT and BPRT larger under the logit link.

Because the reported post-treatment ATT numerators are positive, their lower-bound interpretation under possible backlash suggests that the qualitative conclusion is not driven solely by the no-backlash condition. \Cref{app:general-link-covariates} reports a covariate-adjusted illustration for this application, using province gross regional product (GRP) per capita as a predetermined covariate.

\section{Conclusion}\label{section:conclusion}

This paper develops forward and backward persuasion rates for staggered DID designs with binary outcomes. The rates measure treatment-induced responses among units who would not otherwise respond and the share of observed treated
responses attributable to treatment. Known-link parallel trends identify the counterfactual cell probabilities entering both rates, while event-study aggregation preserves their conditional-probability interpretation by aggregating causal numerators and denominator populations separately. Under no backlash, these probabilities also identify latent response-type shares; without it, they deliver sharp lower bounds. We provide a GMM-based full-information inference procedure using microdata as well as a limited-information procedure when only an ATT and partial denominator information are available. We also consider alternative links to assess sensitivity to the scale on which parallel trends are imposed.

The curriculum application supports the original qualitative finding that exposure to the new curriculum led to more reform-aligned views of China's governance. Our estimates of the forward rates typically exceed the estimated backward rates, consistent with the estimated share of already-persuaded students being larger than the share of never-persuadable students. The event-study forward rates also tend to exceed the rate reported in the original study. The latter difference reflects two departures from the original calculation: we aggregate cohort-time numerators and denominator populations separately rather than rescaling a single DID coefficient, and we exclude provinces treated before 2007 from the comparison sample. Moving from the identity to the logit link changes the magnitudes but not the broad empirical conclusion. More generally, the denominator population, aggregation rule, and probability scale for parallel trends are identifying choices in empirical persuasion analysis.

An important direction for future research is to extend persuasion rates to richer treatment paths. Media DID designs often feature continuous exposure, as in \citet{wang2021media}, or multivalued and nonabsorbing treatments, as in \citet{gentzkow2011effect}. Recent DID work develops methods for continuous treatments \citep{callaway2024continuous} and for complex multivalued, nonabsorbing treatment paths \citep{dechaisemartin2025complex}. Extending persuasion rates to these settings would require defining counterfactual response probabilities and denominator populations for alternative treatment paths. Doing so would make persuasion-rate analysis applicable to media and policy interventions with continuous or changing exposure.

\bibliographystyle{ecta-fullname}

{\small
\bibliography{persuasion2}
}
\newpage

\begin{appendix}

\pagenumbering{roman}
\setcounter{page}{1}

\renewcommand{\thesection}{S-\arabic{section}}
\setcounter{section}{0}

	\begin{center}
            \Large{Online Appendices to ``Learning the Effect of Persuasion via Difference-in-Differences''}
        \end{center}
        \bigskip
        \begin{center}
            \begin{tabular}{ccc}
                Sung Jae Jun & \qquad & Sokbae Lee  \\
                Penn State University & \qquad  & Columbia University
                \end{tabular}
        \end{center}
        \bigskip

\section{Proofs of the Results in the Main Text}\label{app:proofs}

\subsection{\texorpdfstring{Proof of \Cref{thm:id-cell}}{Proof of the Identification Theorem}}\label{app:proof-thm-id-cell}

\begin{proof}
Fix a treated cohort $s$ and period $t\geq s$. By \Cref{ass:stagger-common-trend},
\begin{multline*}
\Lambda\bigl[\tau(s,t)\bigr]
=
\Lambda\bigl[\Pr\{Y_{i,s-1}(\infty)=1\mid S_i=s\}\bigr]
\\
+\Lambda\bigl[\Pr\{Y_{it}(\infty)=1\mid S_i=\infty\}\bigr]
-\Lambda\bigl[\Pr\{Y_{i,s-1}(\infty)=1\mid S_i=\infty\}\bigr].
\end{multline*}
By \Cref{ass:stagger-did}, two observed-outcome equalities apply. First, for units in treated cohort $s$, no anticipation implies $Y_{i,s-1}=Y_{i,s-1}(\infty)$ in the pre-treatment period $s-1$. Second, for units with $S_i=\infty$, because they are never treated, $Y_{i,s-1}=Y_{i,s-1}(\infty)$ and $Y_{it}=Y_{it}(\infty)$ in the two comparison periods $s-1$ and $t$. Therefore, the right-hand side is the argument of $\Lambda^{-1}$ in \eqref{eq:psi-lambda}. Since $\Lambda$ is strictly increasing, it is invertible on its range, and therefore $\tau(s,t)=\psi_{s,t}(\Lambda)$.

Under \Cref{ass:mtr}, \eqref{eq:tp-and-att} shows that the common numerator of FPRT and BPRT is the cohort-time ATT,
$\mu_{s,t}-\tau(s,t)$. Substituting $\tau(s,t)=\psi_{s,t}(\Lambda)$ into \eqref{eq:f,b,att} gives the cohort-time formulas.

For event time $j$, aggregate the cohort-time numerators and denominators in the definitions of FES and BES over all eligible cohorts $s\in\mathcal S_j$. The FES numerator is
\[
\sum_{s\in\mathcal S_j}\pi_s
\{\mu_{s,s+j}-\psi_{s,s+j}(\Lambda)\},
\]
and its denominator is
\[
\sum_{s\in\mathcal S_j}\pi_s\{1-\psi_{s,s+j}(\Lambda)\}.
\]
The BES numerator is the same, while its denominator is $\sum_{s\in\mathcal S_j}\pi_s\mu_{s,s+j}$. These are exactly the stated event-study formulas. The positive-probability condition in \Cref{ass:stagger-did} ensures that the cohort-time denominators are positive. For each eligible cohort $s$ at event time $j$, set $t=s+j$. For FPRT,
$\Pr\{Y_{i,s+j}(\infty)=0,\ S_i=s\}>0$ implies $1-\tau(s,s+j)>0$, and hence $1-\psi_{s,s+j}(\Lambda)>0$. For BPRT,
$\Pr\{Y_{i,s+j}=1,\ S_i=s\}>0$ implies $\mu_{s,s+j}>0$. Summing the corresponding positive terms over eligible cohorts gives positive event-study denominators.
\end{proof}

\subsection{\texorpdfstring{Proof of \Cref{cor:types}}{Proof of the Persuasion-Type Identification Corollary}}\label{app:proof-cor-types}

\begin{proof}
Fix a treated cohort $s$ and period $t\geq s$. For units with $S_i=s$ we have $Y_{it}=Y_{it}(s)$, so $\Pr\{Y_{it}(s)=1\mid S_i=s\}=\mu_{s,t}$. Under \Cref{ass:mtr}, $\Pr\{Y_{it}(s)\geq Y_{it}(\infty)\mid S_i=s\}=1$; hence the event $\{Y_{it}(s)=0,\,Y_{it}(\infty)=1\}$ is null given $S_i=s$, so $\TC(s,t)=0$. The same monotonicity gives, up to null sets given $S_i=s$, $\{Y_{it}(\infty)=1\}\subseteq\{Y_{it}(s)=1\}$ and $\{Y_{it}(s)=0\}\subseteq\{Y_{it}(\infty)=0\}$, so that
\[
\AP(s,t)=\Pr\{Y_{it}(\infty)=1\mid S_i=s\}=\tau(s,t),
\quad
\NP(s,t)=\Pr\{Y_{it}(s)=0\mid S_i=s\}=1-\mu_{s,t}.
\]
By \Cref{thm:id-cell}, $\tau(s,t)=\psi_{s,t}(\Lambda)$, so $\AP(s,t)=\psi_{s,t}(\Lambda)$. Since the four type shares sum to one, $\TP(s,t)=1-\AP(s,t)-\NP(s,t)-\TC(s,t)=\mu_{s,t}-\psi_{s,t}(\Lambda)$.
\end{proof}

\subsection{\texorpdfstring{Proof of \Cref{cor:2x2-cantoni}}{Proof of the Two-Period Identity-Link Benchmark}}\label{app:proof-cor-2x2-cantoni}

\begin{proof}
Specialize \Cref{thm:id-cell} to the identity link $\Lambda_{\mathrm{id}}(p)=p$ and to $t=s$. From \eqref{eq:psi-lambda},
\[
\psi_{s,s}(\Lambda_{\mathrm{id}})
=\mu_{s,s-1}+\mu_{\infty,s}-\mu_{\infty,s-1},
\]
so the common numerator of the two rates equals
\[
\mu_{s,s}-\psi_{s,s}(\Lambda_{\mathrm{id}})
=(\mu_{s,s}-\mu_{s,s-1})-(\mu_{\infty,s}-\mu_{\infty,s-1})
=\att(s,s).
\]
Substitute this numerator into the cohort-time formulas of \Cref{thm:id-cell}, and the corollary follows.
\end{proof}

\subsection{\texorpdfstring{Proof of \Cref{thm:id-logit-rate-gap}}{Proof of the Persuasion-Rate Differences Theorem}}\label{app:proof-thm-id-logit-rate-gap}

\begin{proof}
Recall from the definition of $\psi_{s,s+j}(\Lambda_{\mathrm{logit}})$ that
\[
\Lambda_{\mathrm{logit}}\{ \psi_{s,s+j}(\Lambda_{\mathrm{logit}}) \}
=
\Lambda_{\mathrm{logit}}(\mu_{s,s-1})
+
\Lambda_{\mathrm{logit}}(\mu_{\infty,s+j})
-
\Lambda_{\mathrm{logit}}(\mu_{\infty,s-1}).
\]
Exponentiating both sides yields
\begin{align*}
\frac{\psi_{s,s+j}(\Lambda_{\mathrm{logit}})}
{1-\psi_{s,s+j}(\Lambda_{\mathrm{logit}})}
&=
\frac{\mu_{s,s-1}}{1-\mu_{s,s-1}}
\cdot
\frac{\mu_{\infty,s+j}}{1-\mu_{\infty,s+j}}
\cdot
\frac{1-\mu_{\infty,s-1}}{\mu_{\infty,s-1}}.
\end{align*}
Converting odds to probabilities therefore yields
\[
\psi_{s,s+j}(\Lambda_{\mathrm{logit}})
=
\frac{
\mu_{s,s-1}\mu_{\infty,s+j}(1-\mu_{\infty,s-1})
}{
\mu_{s,s-1}\mu_{\infty,s+j}(1-\mu_{\infty,s-1})
+
(1-\mu_{s,s-1})(1-\mu_{\infty,s+j})\mu_{\infty,s-1}
}.
\]

To verify that
\begin{align*}
\Delta_\psi(s,s+j)
=
\frac{N_\psi(s,s+j)}{D_\psi(s,s+j)},
\end{align*}
we multiply
$\Delta_\psi(s,s+j)$ by $D_\psi(s,s+j)$, yielding
\begin{align*}
D_\psi(s,s+j)\Delta_\psi(s,s+j)
&=
D_\psi(s,s+j)
\{\psi_{s,s+j}(\Lambda_{\mathrm{logit}})-\psi_{s,s+j}(\Lambda_{\mathrm{id}})\}\\
&=
\mu_{s,s-1}\mu_{\infty,s+j}(1-\mu_{\infty,s-1})
-\{\mu_{s,s-1}+\mu_{\infty,s+j}-\mu_{\infty,s-1}\}
D_\psi(s,s+j)\\
&=
\{\mu_{s,s-1}-\mu_{\infty,s-1}\}
\{\mu_{\infty,s+j}-\mu_{\infty,s-1}\}
\{1-\mu_{s,s-1}-\mu_{\infty,s+j}\}\\
&=N_\psi(s,s+j).
\end{align*}
The penultimate equality follows by substituting the definition of $D_\psi(s,s+j)$ and collecting terms.
Since the probabilities entering the logit expression lie in $(0,1)$, $D_\psi(s,s+j)>0$ from its definition. Dividing by $D_\psi(s,s+j)$ gives the stated expression for $\Delta_\psi(s,s+j)$.

For FPRT,
\begin{align*}
\theta^{(F)}_{\mathrm{logit}}(s,s+j) &-\theta^{(F)}_{\mathrm{id}}(s,s+j)
\\
&=
\frac{\mu_{s,s+j}-\psi_{s,s+j}(\Lambda_{\mathrm{logit}})} {1-\psi_{s,s+j}(\Lambda_{\mathrm{logit}})}
-
\frac{\mu_{s,s+j}-\psi_{s,s+j}(\Lambda_{\mathrm{id}})} {1-\psi_{s,s+j}(\Lambda_{\mathrm{id}})}
\\
&=
\frac{\{\mu_{s,s+j}-\psi_{s,s+j}(\Lambda_{\mathrm{logit}})\}\{1-\psi_{s,s+j}(\Lambda_{\mathrm{id}})\} }
{ \{1-\psi_{s,s+j}(\Lambda_{\mathrm{id}})\} \{1-\psi_{s,s+j}(\Lambda_{\mathrm{logit}})\} }
\\
&\qquad\qquad\qquad\qquad\qquad
-
\frac{ \{\mu_{s,s+j}-\psi_{s,s+j}(\Lambda_{\mathrm{id}})\} \{1-\psi_{s,s+j}(\Lambda_{\mathrm{logit}})\} }
 {  \{1-\psi_{s,s+j}(\Lambda_{\mathrm{id}})\}\{1-\psi_{s,s+j}(\Lambda_{\mathrm{logit}})\}  }
\\
&=
\{\psi_{s,s+j}(\Lambda_{\mathrm{id}})-\psi_{s,s+j}(\Lambda_{\mathrm{logit}})\}
\frac{1-\mu_{s,s+j}}{\{1-\psi_{s,s+j}(\Lambda_{\mathrm{id}})\}\{1-\psi_{s,s+j}(\Lambda_{\mathrm{logit}})\}}
\\
&=
-\Delta_\psi(s,s+j) \frac{1-\mu_{s,s+j}} {\{1-\psi_{s,s+j}(\Lambda_{\mathrm{id}})\}\{1-\psi_{s,s+j}(\Lambda_{\mathrm{logit}})\}}.
\end{align*}
For BPRT,
\begin{align*}
\theta^{(B)}_{\mathrm{logit}}(s,s+j)-\theta^{(B)}_{\mathrm{id}}(s,s+j)
&=
\frac{\mu_{s,s+j}-\psi_{s,s+j}(\Lambda_{\mathrm{logit}})}
{\mu_{s,s+j}}
-
\frac{\mu_{s,s+j}-\psi_{s,s+j}(\Lambda_{\mathrm{id}})}
{\mu_{s,s+j}}\\
&=
\frac{\psi_{s,s+j}(\Lambda_{\mathrm{id}})-\psi_{s,s+j}(\Lambda_{\mathrm{logit}})}
{\mu_{s,s+j}}\\
&=
-\frac{\Delta_\psi(s,s+j)}{\mu_{s,s+j}}.
\end{align*}
\end{proof}

\subsection{Additional Details on Limited-Information Inference}
\label{app:limited-info-details}
When $\alpha_q>0$, one convenient choice for $[\underline q_{\ES}(j), \overline q_{\ES}(j)]$ is the normal-approximation confidence interval for $q_{\ES}(j)$ intersected with its known support:
\[
\left[
\widehat q_{\ES}(j)
-z_{1-\alpha_q/2}\widehat{\mathrm{se}}_{q,\ES}(j),
\ \widehat q_{\ES}(j)
+z_{1-\alpha_q/2}\widehat{\mathrm{se}}_{q,\ES}(j)
\right]
\cap[0,1].
\]
Here $\widehat{\mathrm{se}}_{q,\ES}(j)$ is obtained by applying the delta
method to the estimated treated-cell means and weights, with clustering at
the level of treatment assignment. This interval is used in
\Cref{sec:example:limited}. More generally, any interval with asymptotic
coverage at least $1-\alpha_q$ for $q_{\ES}(j)$ may be used. If a range for
$q_{\ES}(j)$ is imposed as deterministic auxiliary information, one instead
sets $\alpha_q=0$.

Intersection with $[0,1]$ can yield $\overline q_{\ES}(j)=1$ in finite samples, in which case $\widehat U^B_{\ES}(j)$ has a zero denominator. On this event, $\widehat U^B_{\ES}(j)$ may be assigned any fixed value. Under the conditions of \Cref{thm:limited-info-ci}, the probability of this event is $o(1)$, so this assignment does not affect asymptotic coverage.

\subsection{\texorpdfstring{Proof of \Cref{thm:limited-info-ci}}{Proof of the Limited-Information Confidence Intervals Theorem}}\label{app:proof-thm-limited-info-ci}
\begin{proof}
For ease of notation, write
\[
\begin{aligned}
\widehat a&:=\widehat{\att}_{\ES}(j), &
a_0&:=\att_{\ES}(j), &
q_0&:=q_{\ES}(j),\\
\overline q&:=\overline q_{\ES}(j), &
\underline q&:=\underline q_{\ES}(j), &
\widehat s &:= \widehat{\mathrm{se}}_{\att,\ES}(j).
\end{aligned}
\]
On events where $\widehat s=0$, set all ratios involving $\widehat s$ in this proof equal to zero; these events have probability $o(1)$. The forward rate is well-defined because $a_0+q_0>0$ is assumed. For the backward rate, the definition of $q_0$, the nonnegativity and normalization of the weights, and the observation rule in \Cref{ass:stagger-did} give
\[
1-q_0
=
\sum_{s\in\mathcal S^{\mathrm{LI}}_j}
\omega_{s,j}\mu_{s,s+j}
=
\sum_{s\in\mathcal S^{\mathrm{LI}}_j}
\omega_{s,j}
\Pr\{Y_{i,s+j}(s)=1\mid S_i=s\}
>0,
\]
where the strict inequality follows from the positivity requirement in
\Cref{ass:stagger-did}. Thus both population rates are well-defined.

By the bounded-away-from-zero, ordering, positivity, and
consistency conditions in the theorem, there is a constant $c>0$ such that
the event
\[
\mathcal A_n
:=
\Bigl\{
0\leq\underline q\leq\overline q<1,\quad
a_0+\underline q \geq c,\quad
1-\overline q \geq c,\quad
\widehat s > 0,\quad
|\widehat a-a_0|\leq c/2
\Bigr\}
\]
satisfies $\Pr(\mathcal A_n)=1-o(1)$. All endpoint arguments below are made
on $\mathcal A_n$. Since $\Pr(\mathcal A_n^c)=o(1)$, restricting attention
to this event affects the probability bounds only through $o(1)$ terms. In
particular, for $q\in\{\underline q,\overline q\}$,
\[
a_0+q\geq c,
\qquad
\widehat a+q\geq c/2,
\qquad
1-q\geq c.
\]
Moreover, if $\underline q=0$, then $a_0\geq c$ and
$\widehat a\geq c/2>0$; if $\overline q=0$, the ordering in
$\mathcal A_n$ implies $\underline q=0$.

We start with the forward case. Let
\[
h_F(a,q) := \frac{a}{a+q}.
\]
For $a\geq 0$, $h_F(a,q)$ is nonincreasing in $q$. Hence, on the event
$q_0\in[\underline q,\overline q]$, we have
\begin{align}\label{eq:hF-monotone}
h_F(a_0,\overline q)
\leq
\theta^{(F)}_{\ES}(j)
\leq
h_F(a_0,\underline q).
\end{align}
Focus on the upper endpoint and consider the event
\[
E_{F,U}
:=
\Bigl\{
\widehat U^F_{\ES}(j)
\geq
h_F(a_0,\underline q)
\Bigr\}.
\]
Below we will show that $\Pr(E_{F,U}^c) \leq \alpha_{\att}/2 + o(1)$.

First,
\begin{equation}\label{eq:split-EFU}
\Pr(E_{F,U}^c )
\leq
\Pr( \mathcal A_n^c )
+
\Pr\bigl( \mathcal A_n \cap \{ \underline{q} = 0 \} \cap E_{F,U}^c \bigr)
+
\Pr\bigl( \mathcal A_n \cap \{ \underline{q} > 0 \} \cap E_{F,U}^c \bigr),
\end{equation}
where the first term on the right-hand side is $o(1)$. On $\mathcal A_n$, if $\underline q=0$, then $a_0>0$ and $\widehat a>0$, and hence $\widehat U^F_{\ES}(j)=h_F(\widehat a,0)=h_F(a_0,0)=1$; thus $E_{F,U}$ trivially holds, and the second term on the right-hand side is also negligible. Therefore, we focus on the third probability on the right-hand side.

Assume that $\mathcal A_n$ and $\underline q>0$ hold. By definition,
\[
E_{F,U}
=
\Bigl\{
h_F(\widehat a,\underline q)
+z_{1-\alpha_{\att}/2}\widehat s
H_F(\widehat a,\underline q)
\geq h_F(a_0,\underline q)
\Bigr\},
\]
where
\[
H_F(a,q):=\frac{\partial h_F(a,q)}{\partial a}
=\frac{q}{(a+q)^2}.
\]
Rearranging the defining inequality gives
\[
h_F(\widehat a,\underline q)-h_F(a_0,\underline q)
\geq
-z_{1-\alpha_{\att}/2}\widehat s
H_F(\widehat a,\underline q).
\]
Because $\widehat s>0$ and $\widehat a+\underline q>0$ on $\mathcal A_n$ and $\{\underline q > 0\}$, the quantity $\widehat sH_F(\widehat a,\underline q)$ is positive. Dividing by it does
not reverse the inequality, so $E_{F,U}$ is equivalent to
\[
\frac{h_F(\widehat a,\underline q)-h_F(a_0,\underline q)}
{\widehat sH_F(\widehat a,\underline q)}
\geq
-z_{1-\alpha_{\att}/2}.
\]
The negative sign is therefore a consequence of moving the positive critical-value adjustment to the right-hand side. We dealt with the case $\underline q=0$ separately because $H_F(\widehat a,0)=0$.

For $q\in\{\underline q,\overline q\}$, define on $\mathcal A_n$
\[
T_F(q)
:=
\frac{\widehat a-a_0}{\widehat s}
\frac{\widehat a+q}{a_0+q},
\]
and set $T_F(q)=0$ on $\mathcal A_n^c$. Then, for either $q\in\{\underline q,\overline q\}$, on $\mathcal A_n\cap\{q>0\}$, direct algebra gives the exact identity
\[
\frac{h_F(\widehat a,q)-h_F(a_0,q)}
{\widehat sH_F(\widehat a,q)}
=
T_F(q).
\]
Furthermore, on $\mathcal A_n$,
\begin{equation}\label{eq:error1}
\max_{q\in\{\underline q,\overline q\}}
\left|
\frac{\widehat a+q}{a_0+q}-1
\right|
\leq
\frac{|\widehat a-a_0|}{a_0+\underline q}
\leq
\frac{|\widehat a-a_0|}{c},
\end{equation}
which implies that
\begin{equation}\label{eq:error2}
R_n(q)
:=
T_F(q)
-
\frac{\widehat a - a_0}{\widehat s},
\qquad
q\in\{\underline q,\overline q\},
\end{equation}
and $\max_{q\in\{\underline q,\overline q\}}|R_n(q)| \leq |\widehat a - a_0|^2 / ( c \widehat s)$ on $\mathcal A_n$. Consequently,
\begin{align*}
\Pr\bigl( \mathcal A_n,\ \underline{q}>0,\  E_{F,U}^c \bigr)
&=
\Pr\Bigl\{ \mathcal A_n,\ \underline{q}>0,\ T_F(\underline q) < - z_{1-\alpha_\att /2} \Bigr\}
\\
&\leq
\Pr\Bigg\{\mathcal A_n,\ \underline{q}>0,\ \frac{\widehat a - a_0}{\widehat s}  - \frac{|\widehat a - a_0|^2}{c\widehat s}< - z_{1-\alpha_\att /2}   \Biggr\}
\\
&\leq
\Pr\Bigg\{\frac{\widehat a - a_0}{\widehat s}  - \frac{|\widehat a - a_0|^2}{c\widehat s}< - z_{1-\alpha_\att /2}   \Biggr\}.
\end{align*}
But, we have $(\widehat a - a_0)/\widehat s \convd N(0,1)$ and $|\widehat a - a_0| = o_p(1)$, and hence it follows from Slutsky's theorem that
\begin{align*}
\Pr\Bigg\{\frac{\widehat a - a_0}{\widehat s}  - \frac{|\widehat a - a_0|^2}{c\widehat s}< - z_{1-\alpha_\att /2}   \Biggr\}
&=
\Pr\Bigg\{\frac{\widehat a - a_0}{\widehat s} < - z_{1-\alpha_\att /2}   \Biggr\} +o(1)
\\
&=
\alpha_\att/2 + o(1).
\end{align*}
Combine this with \Cref{eq:split-EFU}, and it follows that
\[
\Pr( E_{F,U}^c) \leq \alpha_{\att}/2 + o(1).
\]

Similarly, consider the lower-endpoint event
\[
E_{F,L}
:=
\Bigl\{
\widehat L^F_{\ES}(j)
\leq
h_F(a_0,\overline q)
\Bigr\}.
\]
Note that
\[
\Pr(E_{F,L}^c)
\leq
\Pr(\mathcal A_n^c)
+
\Pr(\mathcal A_n\cap \{ \overline q = 0\}\cap E_{F,L}^c)
+
\Pr(\mathcal A_n\cap \{ \overline q > 0\}\cap E_{F,L}^c),
\]
where the first term on the right-hand side is $o(1)$. For the second term, $\overline{q} = 0$ and the ordering in $\mathcal A_n$ implies that $\underline q = 0$, and hence $a_0 > 0$ and $\widehat a>0$. Then, it follows that $\widehat L^F_{\ES}(j)=h_F(\widehat a,0)=h_F(a_0,0)=1$; hence $E_{F,L}$ trivially holds, and the second term on the right-hand side is negligible. Finally, when $\mathcal A_n$ and $\overline q >0$ hold, the same rearrangement as in the case of $E_{F,U}$ shows that $E_{F,L}$ can be equivalently expressed as $T_F(\overline q) \leq z_{1-\alpha_{\att}/2}$. Therefore, by using \Cref{eq:error1,eq:error2} as before,
\begin{align*}
\Pr(\mathcal A_n,\ \{ \overline q > 0\},\ E_{F,L}^c)
&=
\Pr(\mathcal A_n,\ \{ \overline q > 0\},\ T_F(\overline q) > z_{1-\alpha_{\att}/2})
\\
&\leq
\Pr\left(\mathcal A_n,\ \{ \overline q > 0\},\ \frac{\widehat a - a_0}{\widehat s}  + \frac{|\widehat a - a_0|^2}{c\widehat s}  > z_{1-\alpha_{\att}/2} \right)
\\
&\leq
\Pr\left(\frac{\widehat a - a_0}{\widehat s}  + \frac{|\widehat a - a_0|^2}{c\widehat s}  > z_{1-\alpha_{\att}/2} \right)
=
\alpha_{\att}/2 + o(1),
\end{align*}
where the last equality is by the asymptotic normality of $(\widehat a - a_0)/\widehat s$ and Slutsky. Therefore, we have shown that
\[
\Pr(E_{F,L}^c)\leq \alpha_{\att}/2 + o(1).
\]

Now define
$E_q:=\{q_0\in[\underline q,\overline q]\}$. Then
\cref{eq:hF-monotone} shows that
$E_q\cap E_{F,L}\cap E_{F,U}$ implies
$\theta^{(F)}_{\ES}(j)\in
[\widehat L^F_{\ES}(j),\widehat U^F_{\ES}(j)]$. Therefore,
\begin{multline*}
\Pr\Bigl\{\theta^{(F)}_{\ES}(j)\notin
[\widehat L^F_{\ES}(j),\widehat U^F_{\ES}(j)]\Bigr\}
\leq
\Pr(E_q^c\cup E_{F,L}^c\cup E_{F,U}^c)
\\
\leq
\Pr(E_q^c)+\Pr(E_{F,L}^c)+\Pr(E_{F,U}^c)
\leq
\alpha_q+\alpha_{\att}/2+\alpha_{\att}/2+o(1)
=\alpha+o(1).
\end{multline*}

For the backward rate, define
\[
h_B(a,q):=\frac{a}{1-q},
\]
which is nondecreasing in $q$ for $a\geq0$. Hence, on $E_q$, we have
\begin{align}\label{eq:hB-monotone}
h_B(a_0,\underline q)
\leq
\theta^{(B)}_{\ES}(j)
\leq
h_B(a_0,\overline q).
\end{align}
Define the endpoint events
\[
E_{B,L}
:=
\Bigl\{
\widehat L^B_{\ES}(j)
\leq
h_B(a_0,\underline q)
\Bigr\},
\qquad
E_{B,U}
:=
\Bigl\{
\widehat U^B_{\ES}(j)
\geq
h_B(a_0,\overline q)
\Bigr\}.
\]
Writing
$H_B(a,q):=\partial h_B(a,q)/\partial a=1/(1-q)$ gives the exact identity
\[
\frac{h_B(\widehat a,q)-h_B(a_0,q)}
{\widehat sH_B(\widehat a,q)}
=
\frac{\widehat a-a_0}{\widehat s},
\]
even when $q$ is random. On $\mathcal A_n$, $\widehat s>0$
and $1-q\geq c$ for $q\in\{\underline q,\overline q\}$, so the standardized
statistic and the displayed ratios at both endpoints are well-defined, with
positive denominators. On exceptional events where $\widehat s=0$, the
standardized statistic is assigned an arbitrary value under the convention
above; these events are contained in $\mathcal A_n^c$ and affect the bounds
below only through $o(1)$ terms. The definitions of the two confidence
limits therefore imply
\[
E_{B,L}
=
\left\{
\frac{\widehat a-a_0}{\widehat s}
\leq z_{1-\alpha_{\att}/2}
\right\},
\qquad
E_{B,U}
=
\left\{
\frac{\widehat a-a_0}{\widehat s}
\geq -z_{1-\alpha_{\att}/2}
\right\}
\]
on $\mathcal A_n$. Thus the assumed standardized asymptotic normality yields
\begin{align*}
\Pr(E_{B,L}^c)
&\leq
\Pr(\mathcal A_n^c)
+\Pr\left\{
\frac{\widehat a-a_0}{\widehat s}
>z_{1-\alpha_{\att}/2}
\right\}
\leq \alpha_{\att}/2+o(1),\\
\Pr(E_{B,U}^c)
&\leq
\Pr(\mathcal A_n^c)
+\Pr\left\{
\frac{\widehat a-a_0}{\widehat s}
<-z_{1-\alpha_{\att}/2}
\right\}
\leq \alpha_{\att}/2+o(1).
\end{align*}
By \cref{eq:hB-monotone},
$E_q\cap E_{B,L}\cap E_{B,U}$ implies
$\theta^{(B)}_{\ES}(j)\in
[\widehat L^B_{\ES}(j),\widehat U^B_{\ES}(j)]$. Consequently,
\begin{multline*}
\Pr\Bigl\{\theta^{(B)}_{\ES}(j)\notin
[\widehat L^B_{\ES}(j),\widehat U^B_{\ES}(j)]\Bigr\}
\\
\leq
\Pr(E_q^c)+\Pr(E_{B,L}^c)+\Pr(E_{B,U}^c)
\leq
\alpha_q+\alpha_{\att}/2+\alpha_{\att}/2+o(1)
=\alpha+o(1),
\end{multline*}
which proves the result for the backward rate.
\end{proof}

\section{Covariate-Adjusted Cell-Probability Estimation}\label{app:general-link-covariates}

This appendix extends the cell-probability framework of \Cref{sec:est:inf} to allow adjustment for baseline or pre-treatment covariates. The extension keeps the two features emphasized in the main text: the data may be panel data, repeated cross-sections, or a single-survey cohort design, and the parallel-trends restriction may be imposed after applying a known link function $\Lambda$.

\subsection{Conditional Cell Means}

Let $X_i$ denote baseline or pre-treatment covariates observed for the cells used in the analysis. For $a\in\{1,\ldots,T,\infty\}$ and $t\in\{0,\ldots,T\}$, define
\[
\mu_{a,t}(x)
:=
\Pr\{Y_{i,at}^{\mathrm{obs}}=1\mid C_{i,at}=1,\ X_i=x\}.
\]
This is the conditional counterpart of the unconditional cell mean $\mu_{a,t}$ in the main text. The covariate-adjusted restriction imposes the link-scale parallel-trends condition of \Cref{ass:stagger-common-trend} conditional on $X_i$: for treated cohort $s$, post-treatment period $t\geq s$, and $X_i = x$,
\begin{multline*}
\Lambda\bigl[\Pr\{Y_{it}(\infty)=1\mid S_i=s,\ X_i=x\}\bigr]
-
\Lambda\bigl[\Pr\{Y_{i,s-1}(\infty)=1\mid S_i=s,\ X_i=x\}\bigr]
\\
=
\Lambda\bigl[\Pr\{Y_{it}(\infty)=1\mid S_i=\infty,\ X_i=x\}\bigr]
-
\Lambda\bigl[\Pr\{Y_{i,s-1}(\infty)=1\mid S_i=\infty,\ X_i=x\}\bigr],
\end{multline*}
where the probabilities lie in the domain of $\Lambda$. Together with no anticipation, this restriction identifies the conditional untreated counterfactual for the treated cohort:
\begin{equation}\label{eq:cond-psi-lambda}
\psi_{s,s+j}(x;\Lambda)
:=
\Lambda^{-1}\left\{
\Lambda\{\mu_{s,s-1}(x)\}
+
\Lambda\{\mu_{\infty,s+j}(x)\}
-
\Lambda\{\mu_{\infty,s-1}(x)\}
\right\}.
\end{equation}
\Cref{eq:cond-psi-lambda} is the conditional analog of \eqref{eq:psi-lambda}.

\subsection{Covariate-Adjusted Event-Study Rates}

Let $H_{s,j}$ denote the distribution of $X_i$ in the treated target cell $C_{i,s,s+j}=1$. In panel data $H_{s,j}$ is the covariate distribution of cohort $s$. In repeated cross-sections or a single-survey cohort design, it is the covariate distribution in the observed target cohort-time cell. This target distribution is part of the estimand: changing $H_{s,j}$ affects the population rate being summarized as the conditional quantities are averaged over the treated target covariate distribution. This target-cell averaging preserves the numerator-denominator structure of \Cref{thm:id-cell}:
\begin{align}
\theta^{(F)}_{\ES,H}(j;\Lambda)
&=
\frac{
\sum_{s\in\mathcal S_j}\pi_s
\int
\{\mu_{s,s+j}(x)-\psi_{s,s+j}(x;\Lambda)\}
\,dH_{s,j}(x)
}{
\sum_{s\in\mathcal S_j}\pi_s
\int
\{1-\psi_{s,s+j}(x;\Lambda)\}
\,dH_{s,j}(x)
},
\label{eq:covariate-fes}
\\
\theta^{(B)}_{\ES,H}(j;\Lambda)
&=
\frac{
\sum_{s\in\mathcal S_j}\pi_s
\int
\{\mu_{s,s+j}(x)-\psi_{s,s+j}(x;\Lambda)\}
\,dH_{s,j}(x)
}{
\sum_{s\in\mathcal S_j}\pi_s
\int
\mu_{s,s+j}(x)
\,dH_{s,j}(x)
}.
\label{eq:covariate-bes}
\end{align}
The subscript $H$ distinguishes these covariate-standardized rates from the
unconditional event-study rates in the main text. For the identity link, the conditional quantities enter linearly before being averaged over the treated target distribution. For a nonlinear link, the link transformation is applied conditionally through \eqref{eq:cond-psi-lambda} before the averaging step; this generally differs from applying the nonlinear link to unconditional cell means. As in the main text, the persuasion-rate interpretation also uses no backlash on the treated target support.

\subsection{Estimation}

Estimation proceeds by replacing each object in the covariate-adjusted formulas
with its sample analog. First estimate each required conditional mean $\mu_{a,t}(x)$ using observations in the corresponding cell $C_{i,at}=1$. Then form $\widehat\psi_{s,s+j}(x;\Lambda)$ by substituting the fitted conditional means into \eqref{eq:cond-psi-lambda}. The integrals with respect to $H_{s,j}$ are evaluated as empirical averages over observations with $C_{i,s,s+j}=1$, while $\pi_s$ and $\mathcal S_j$ are replaced by the treated-cohort shares and eligible sets used in \Cref{sec:est:inf}.

With discrete covariates, this procedure is the fully interacted finite-cell estimator. With continuous covariates, the same plug-in formula can be used after fitting fixed finite-dimensional conditional-mean models. For example, one may estimate separate cell-specific binary-response models. The sample eligible set follows the rule in \Cref{sec:est:inf}, applied to the fitted conditional quantities on the empirical target support: the fitted conditional means must lie in the domain of the chosen link, each constructed $\widehat\psi_{s,s+j}(x;\Lambda)$ must lie in $[0,1]$, and the standardized cohort-time denominators must be strictly positive.

This construction requires the treated target distribution to be supported where each conditional mean entering the counterfactual is identified. With fixed-dimensional parametric first steps, consistency additionally requires correct specification of the corresponding conditional-mean models.

\subsection{Clustered Inference}

For saturated finite cells and fixed finite-dimensional first steps, inference
can be carried out by the standard clustered GMM delta method. Fix a horizon $j$. Form the union, over eligible cohorts $s$, of the four cells entering each
counterfactual: $(s,s+j)$, $(s,s-1)$, $(\infty,s+j)$, and  $(\infty,s-1)$. Stack the score equation once for each distinct cohort-period cell $(a,t)$ in this union. Write the corresponding conditional-mean model as $m_{a,t}(x;\beta_{a,t})$. In a numerical illustration in \Cref{sec:grp-example}, for example, we use the logit first stage
\[
m_{a,t}(x;\beta_{a,t})
=
\ell\{Z(x)^\top\beta_{a,t}\},
\qquad
\ell(u)=\frac{\exp(u)}{1+\exp(u)},
\qquad
Z(x)=(1,\log \mathrm{GRP})^\top,
\]
with score equation
\[
\sum_i
C_{i,a,t}\,
Z(X_i)\{Y_{i,a,t}^{\mathrm{obs}}-m_{a,t}(X_i;\widehat\beta_{a,t})\}
=0.
\]
The standardized target averages are included in the same finite-dimensional
system by first defining the empirical target distribution. Let $\widehat H_{s,j}$ be the empirical distribution of $X_i$ among observations in the treated target cell $C_{i,s,s+j}=1$:
\[
\widehat H_{s,j}(A)
=
\frac{\sum_i C_{i,s,s+j}\one\{X_i\in A\}}
{\sum_i C_{i,s,s+j}}.
\]
The corresponding averages over this treated target distribution are
\[
\widehat\mu^{H}_{s,s+j}
=
\int m_{s,s+j}(x;\widehat\beta_{s,s+j})\,d\widehat H_{s,j}(x)
=
\frac{\sum_i C_{i,s,s+j}
m_{s,s+j}(X_i;\widehat\beta_{s,s+j})}
{\sum_i C_{i,s,s+j}},
\]
and
\[
\widehat\psi^{H}_{s,s+j}(\Lambda)
=
\int \widehat\psi_{s,s+j}(x;\Lambda)\,d\widehat H_{s,j}(x)
=
\frac{\sum_i C_{i,s,s+j}
\widehat\psi_{s,s+j}(X_i;\Lambda)}
{\sum_i C_{i,s,s+j}},
\]
where $\widehat\psi_{s,s+j}(x;\Lambda)$ substitutes the relevant fitted conditional means into \eqref{eq:cond-psi-lambda}. Here, the superscript $H$ in $\widehat\mu^{H}_{s,s+j}$ and $\widehat\psi^{H}_{s,s+j}(\Lambda)$ denotes integration with respect to this empirical target distribution. Equivalently, $\widehat\mu^{H}_{s,s+j}$ and $\widehat\psi^{H}_{s,s+j}(\Lambda)$ can be included in the stacked system through the unnormalized moments
\[
\sum_i
C_{i,s,s+j}
\{m_{s,s+j}(X_i;\widehat\beta_{s,s+j})
-\widehat\mu^{H}_{s,s+j}\}
=0
\]
and
\[
\sum_i
C_{i,s,s+j}
\{\widehat\psi_{s,s+j}(X_i;\Lambda)
-\widehat\psi^{H}_{s,s+j}(\Lambda)\}
=0.
\]
Cohort shares are included
through the moments
\[
\sum_i\{\one(S_i=s)-\widehat\pi_s\}=0.
\]
Let $\widehat\alpha_j$ collect the first-step parameters, target averages, and cohort shares.
Let $\widehat V_{\alpha,j}$ be the cluster-robust sandwich covariance matrix of $\widehat\alpha_j$ from the stacked moments, with moment contributions summed within clusters.
The reported FES and BES standard errors are computed one at a time from this same clustered covariance matrix. These plug-in ratios estimate $\theta^{(F)}_{\ES,H}(j;\Lambda)$ and $\theta^{(B)}_{\ES,H}(j;\Lambda)$. Let $h^F_j(\widehat\alpha_j;\Lambda)$ denote the plug-in ratio in \eqref{eq:covariate-fes}, and let $h^B_j(\widehat\alpha_j;\Lambda)$ denote the plug-in ratio in \eqref{eq:covariate-bes}. Then, for $r\in\{F,B\}$,
\[
\widehat V_{\theta^r,j}
=
\nabla h^r_j(\widehat\alpha_j;\Lambda)^\top
\widehat V_{\alpha,j}
\nabla h^r_j(\widehat\alpha_j;\Lambda).
\]
Because $\widehat V_{\alpha,j}$ estimates $\operatorname{Var}(\widehat\alpha_j)$, the quadratic form $\widehat V_{\theta^r,j}$ is the scalar delta-method variance estimator for $h^r_j(\widehat\alpha_j;\Lambda)$. The reported standard error is therefore $\widehat V_{\theta^r,j}^{1/2}$.

This construction nests the no-covariate estimator used in the main empirical
section. If there are no covariates and $m_{a,t}(x;\beta_{a,t})=\beta_{a,t}$,
the cell-specific score equations reduce to the cell-mean moments in \Cref{sec:est:inf}; the target averages collapse to the corresponding cell means; and FES and BES are evaluated by separate scalar delta-method calculations from the same clustered GMM covariance matrix, exactly as in \Cref{section:example2}.

This argument covers saturated finite cells and fixed-dimensional conditional mean models. Nonparametric or high-dimensional first steps require separate justification.

\subsection{Illustration in the Curriculum Persuasion Example}\label{sec:grp-example}

As a compact illustration, we apply the covariate-adjusted plug-in estimator to the curriculum application studied in \Cref{section:example2}. The outcome is the median-binarized ``Trust: central government'' indicator, and the horizon is $j=0$. The covariate is log province gross regional product (GRP) per capita in 2003, which is predetermined relative to the 2004--2010 curriculum rollout. For each required cohort-time cell, we estimate a separate logit model for the conditional cell mean, linear in log province GRP per capita, and average the fitted quantities over the treated post-treatment target cell. The same cell-specific logit first stage is used in both panels to keep fitted probabilities between zero and one; panels A and B differ only in the link used to construct the counterfactual. The covariate-adjusted estimates are larger than the no-covariate estimates under both links. Under the identity link, FES rises from $0.460$ to $0.579$ and BES rises from $0.187$ to $0.302$. Under the logit link, FES rises from $0.548$ to $0.636$ and BES rises from $0.266$ to $0.384$.

\begin{table}[htbp]
\centering
\caption{Covariate-Adjusted Illustration for Trust in Central Government\label{tab:covariate-curriculum-illustration}}
\begin{tabular}{lcccc}
\hline
Method & FES & s.e. & BES & s.e. \\
\hline
\multicolumn{5}{l}{Panel A: Identity link} \\
No covariates & 0.460 & 0.095 & 0.187 & 0.062 \\
Log GRP & 0.579 & 0.135 & 0.302 & 0.166 \\
\hline
\multicolumn{5}{l}{Panel B: Logit link} \\
No covariates & 0.548 & 0.097 & 0.266 & 0.097 \\
Log GRP & 0.636 & 0.102 & 0.384 & 0.169 \\
\hline
\end{tabular}

\vspace{0.5em}
\begin{minipage}{0.92\textwidth}
\footnotesize
Notes: The outcome is the median-binarized ``Trust: central government'' indicator from the curriculum application. The horizon is $j=0$. Rows labeled ``Log GRP'' use cell-specific logit first stages that are linear in log province gross regional product (GRP) per capita. Standard errors are province-clustered delta-method standard errors based on the finite-dimensional first-step models.
\end{minipage}
\end{table}

\end{appendix}

\end{document}